\documentclass[aps,prd,twocolumn,nofootinbib,superscriptaddress]{revtex4}
\usepackage[]{hyperref}
\usepackage{graphicx}
\def\lesssim{\mathrel{\mathpalette\vereq<}}
\def\gtrsim{\mathrel{\mathpalette\vereq>}}

\newcommand{\be}{\begin{equation}}
\newcommand{\ee}{\end{equation}}
\newcommand{\bea}{\begin{eqnarray}}
\newcommand{\eea}{\end{eqnarray}}

\begin{document}

\pagestyle{plain}

\title{Effective Theory of a Light Dilaton}

\author{Zackaria Chacko}
\author{Rashmish K. Mishra}
\affiliation{Maryland Center for Fundamental Physics, Department of Physics, 
University of Maryland, College Park, MD, 20742}


\begin{abstract}
{
We consider scenarios where strong conformal dynamics constitutes the 
ultraviolet completion of the physics that drives electroweak symmetry 
breaking. We show that in theories where the operator responsible for 
the breaking of conformal symmetry is close to marginal at the breaking 
scale, the dilaton mass can naturally lie below the scale of the strong 
dynamics. However, in general this condition is not satisfied in the 
scenarios of interest for electroweak symmetry breaking, and so the 
presence of a light dilaton in these theories is associated with mild 
tuning. We construct the effective theory of the light dilaton in this 
framework, and determine the form of its couplings to Standard Model 
states. We show that corrections to the form of the dilaton interactions 
arising from conformal symmetry violating effects are suppressed by the 
square of the ratio of the dilaton mass to the strong coupling scale, 
and are under good theoretical control. These corrections are generally 
subleading, except in the case of dilaton couplings to marginal 
operators, when symmetry violating effects can sometimes dominate. We 
investigate the phenomenological implications of these results for 
models of technicolor, and for models of the Higgs as a 
pseudo-Nambu-Goldstone boson, that involve strong conformal dynamics in 
the ultraviolet.
}
\end{abstract}

\pacs{} \maketitle


\section{Introduction}

Although the Standard Model (SM) has experienced more than 30 years of 
experimental successes, the nature of the dynamics that underlies the 
electroweak phase transition has remained a mystery. Although several 
promising theories have been put forward, each faces its own challenges 
and conclusive experimental evidence that could settle the issue one way 
or the other has been lacking. The recent discovery of a new particle 
with mass close to 125 GeV and properties similar to that of the SM 
Higgs boson~\cite{HiggsJuly4CMS, HiggsJuly4ATLAS} will help resolve this 
issue. The simplest possibility, and the one favored by precision 
electroweak tests, is that the new particle is indeed the SM Higgs. In 
this case the remaining challenge is to explain the stability of the 
weak scale under radiative corrections, the `hierarchy problem'. If, 
however, electroweak symmetry is broken by strong dynamics, as in 
technicolor models~\cite{Susskind:1978ms, Weinberg:1975gm}, for a review 
see~\cite{Hill:2002ap}, there is no immediate understanding of the 
origin of the new particle, or an explanation of why its properties are 
similar to that of the SM Higgs. The simplest technicolor models remain 
disfavored by precision tests~\cite{Holdom:1990tc, Golden:1990ig, 
Peskin:1991sw}, and there is no immediate understanding of the absence 
of the new contributions to flavor changing neutral currents that are 
expected to be generated by the mechanism that gives the SM quarks and 
leptons their masses.

There have long been good reasons to think that strong conformal 
dynamics may play a role in electroweak symmetry breaking, irrespective 
of the existence of a light Higgs. In theories of technicolor, if the 
strong dynamics that breaks electroweak symmetry is conformal in the 
ultraviolet, the operators that give rise to the fermion masses can have 
a large anomalous dimension. This framework is used in conformal 
technicolor models~\cite{hep-ph/0409274} to generate a natural 
separation of the flavor scale from the electroweak scale, allowing the 
experimental limits on flavor violation to be satisfied. (This approach 
to the flavor problem was first proposed in the context of walking 
technicolor~\cite{Holdom:1984sk, Appelquist:1986an, Yamawaki:1985zg, 
Appelquist:1986tr}, which is closely related to conformal technicolor.) 
In theories with a light Higgs, one class of promising solutions to the 
hierarchy problem are those where the SM Higgs emerges as the 
pseudo-Nambu-Goldstone boson of a global symmetry that is broken by 
strong dynamics~\cite{Georgi:1974yw, Kaplan:1983fs, Kaplan:1983sm}. This 
class of theories includes little Higgs 
models~\cite{ArkaniHamed:2001nc, ArkaniHamed:2002qy, Gregoire:2002ra}, 
and twin Higgs models~\cite{Chacko:2005pe, Chacko:2005un}. If the 
ultraviolet physics involves strong conformal dynamics, the flavor scale 
can again be separated from the electroweak scale, allowing new 
contributions to flavor violating processes to be small enough to 
satisfy the existing constraints.

In theories where an exact conformal symmetry is spontaneously broken, 
the low energy effective theory below the breaking scale contains a 
massless scalar, the dilaton, which may be thought of as the 
Nambu-Goldstone boson (NGB) associated with the breaking of conformal 
symmetry~\cite{Salam:1970qk, Isham:1971dv, Zuminolectures, 
Ellis:1970yd}. The form of the dilaton couplings is fixed by the 
requirement that conformal symmetry be realized nonlinearly, and so this 
framework is extremely predictive. Several authors have studied the 
couplings of a light dilaton in the context of theories of electroweak 
symmetry breaking~\cite{hep-th/0012248, arXiv:0708.1463, 
arXiv:1002.1721}. Remarkably, the interactions of a dilaton with the SM 
fields are very similar to those of the SM Higgs~\cite{arXiv:0708.1463}. 
This can be traced to the fact that at the classical level the SM has an 
approximate conformal symmetry which is spontaneously broken by the VEV 
of the Higgs, so that the Higgs can be understood as a dilaton in this 
limit. However, in the class of theories of interest for electroweak 
symmetry breaking, conformal symmetry is expected to be explicitly 
violated by operators that grow in the infrared to become strong at the 
breaking scale. Therefore, in general, there is no reason to expect a 
light dilaton in the low energy effective theory.

In this paper we consider scenarios where strong conformal dynamics 
constitutes the ultraviolet completion of the physics that drives 
electroweak symmetry breaking. Following the framework outlined 
in~\cite{RattazziPlanck2010}, we show that in theories where the 
operator responsible for breaking conformal symmetry is marginal at 
the breaking scale, the dilaton mass can naturally lie below the scale 
of the strong dynamics.{\footnote{A closely related result has been 
obtained in the context of walking gauge theories using current 
algebraic methods~\cite{arXiv:1006.4375}.}} However, in general this 
condition is not satisfied by the theories of interest for electroweak 
symmetry breaking, and so the presence of a light dilaton in these 
theories is associated with mild tuning. We construct the effective 
theory of the light dilaton in this framework, and determine the form of 
its couplings to SM states. We show that corrections to the form of the 
dilaton interactions arising from conformal symmetry violating effects 
are suppressed by the square of the ratio of the dilaton mass to the 
strong coupling scale, and are under good theoretical control. These 
corrections are subleading, except in the case of dilaton couplings to 
marginal operators, when they can sometimes dominate.

These results have important implications for our understanding of 
electroweak symmetry breaking. One possibility is that the new particle 
that has been observed close to 125 GeV is not the SM Higgs, but instead 
a dilaton that emerges from a strongly interacting conformal sector that 
breaks electroweak symmetry dynamically. In fact, several papers that 
interpret the 125 GeV resonance as a dilaton have already appeared in 
the literature~\cite{Cheung:2011nv, Matsuzaki:2012gd, 
Grzadkowski:2012ng, Elander:2012fk}. In such a scenario an understanding 
of the general form of the dilaton couplings, including conformal 
symmetry violating effects, is crucial to distinguishing it from the SM 
Higgs~\cite{Chacko:2012vm}. Another possibility is that the new particle which has been 
observed is indeed the SM Higgs, which arises as the 
pseudo-Nambu-Goldstone boson (pNGB) of an approximate global symmetry 
that is broken by strong conformal dynamics. Our analysis shows that in 
such a scenario, there may be an additional light scalar in the low 
energy effective theory beyond the SM Higgs whose couplings to the SM 
fields can be predicted.
 
The AdS/CFT correspondence~\cite{Maldacena:1997re, Witten:1998qj, 
Verlinde:1999fy, Verlinde:1999xm} can be used to relate Randall-Sundrum 
models in warped extra dimensions~\cite{Randall:1999ee} to strongly 
coupled conformal field theories in the large $N$ limit. In this way, 
extra dimensional realizations of technicolor~\cite{Csaki:2003zu} and of 
the Higgs as a pNGB~\cite{Contino:2003ve} have been obtained. In the 
correspondence, the radion in the Randall-Sundrum model is identified 
with the dilaton~\cite{hep-th/0012248}. Radion stabilization using the 
Goldberger-Wise mechanism~\cite{Goldberger:1999uk} can be understood as 
a stable minimum for the dilaton potential being generated by effects 
which explicitly violate conformal symmetry~\cite{hep-th/0012248}. 
Several authors have studied the couplings of the radion in 
Randall-Sundrum models, both in the case when the SM fields are 
localized to a brane~\cite{Goldberger:1999uk, Csaki:1999mp, 
Giudice:2000av, Csaki:2000zn} and in the case when they are in the 
bulk~\cite{Rizzo:2002pq, arXiv:0705.3844}. We find excellent agreement 
between these results and ours in the regime when the large $N$ 
approximation is valid.

\section{Effective Theory of a Dilaton}

In this section we construct the effective theory for the dilaton, 
incorporating conformal symmetry violating effects, and show that if the 
operator that breaks conformal symmetry is marginal at the breaking 
scale, the dilaton can naturally be light.

The fifteen parameter conformal group extends the ten parameter Poincare 
group to include scale transformations and special conformal 
transformations. While it has long been conjectured that any Poincare 
invariant, unitary theory that realizes scale invariance linearly will 
also respect conformal symmetry~\cite{Polchinski:1987dy}, there exists 
no complete proof. The validity of this conjecture has been the subject 
of considerable interest in the recent literature 
~\cite{Fortin:2011ks},~\cite{Luty:2012ww},~\cite{Nakayama:2012nd}.
   
Consider a theory where conformal invariance is spontaneously broken. 
Then the low energy effective theory contains a dilaton field 
$\sigma(x)$, which can be thought of as the NGB associated with the 
breaking of scale invariance~\cite{Salam:1970qk, Isham:1971dv, 
Zuminolectures, Ellis:1970yd}. The additional four NGBs associated with 
the breaking of the special conformal symmetry can be identified with 
the derivatives of the dilaton, rather than as independent propagating 
fields. Below the breaking scale the symmetry is realized non-linearly, 
with the dilaton undergoing a shift $\sigma(x) \rightarrow \sigma'(x') = 
\sigma(x) + \omega f$ under the scale transformation $x^{\mu} 
\rightarrow x'^{\mu} = e^{- \omega} x^{\mu}$. Here $f$ is the scale 
associated with the breaking of conformal symmetry. For the purpose of 
writing interactions of the dilaton it is convenient to define the 
object
 \begin{equation}
\chi(x) = f e^{\sigma(x)/f}
 \end{equation}
which transforms linearly under scale transformations. Specifically, 
under the scale transformation $x^{\mu} \rightarrow x'^{\mu} = 
e^{-\omega} x^{\mu} $, $\chi(x)$ transforms as a conformal compensator
 \begin{equation}
\chi(x) \rightarrow \chi'(x') = e^{\omega} \; \chi(x) \; .
\label{dilatontransformation} 
 \end{equation} 
The low energy effective theory for the dilaton will in general include 
all terms consistent with this transformation, but with some additional 
restrictions and relations among their coefficients from the requirement 
that the theory be invariant not just under scale transformations, but 
under the full conformal group. These restrictions will not affect our 
discussion in any significant way, and so operationally we shall only 
require that the action for $\chi$ be scale invariant.

In writing down the Lagrangian for the dilaton, it is necessary to take 
into account the implicit breaking of conformal invariance associated with 
the regulator. Since the theory in the ultraviolet possesses exact conformal 
invariance, this effect is of course completely spurious. However, it 
has the consequence that the Lagrangian for the dilaton is not 
manifestly scale invariant. It is only at the quantum level, when 
effects of the regulator are incorporated, that conformal invariance is 
realized. This complicates the problem of finding the form of the 
effective theory.

Perhaps the simplest way to incorporate the effect of the regulator is 
to begin in a framework where the renormalization scale $\mu_{\chi}$ is 
itself a function of $\chi$, $\mu_{\chi} = \mu \hat{\chi}$, where 
$\hat{\chi} = \chi/f$. In such a framework, correlation functions can be 
obtained from the effective action, which has exactly the same form as 
in a conventional renormalization scheme, but with $\mu$ replaced by 
$\mu_{\chi}$~\cite{Sundrum:2003yt}. Such a choice of renormalization 
scheme has the advantage that the action for $\chi$ is then manifestly 
scale invariant and therefore easy to write down. Starting from this 
action, the form of the Lagrangian in a more conventional scheme where 
the renormalization scale is independent of $\chi$ can be determined. 
This is the approach we shall follow.

\subsection{Effective Theory in the Limit of Exact Conformal Invariance}

We begin by constructing the effective theory for the dilaton in the 
case when conformal invariance is exact, and effects that explicitly 
violate conformal symmetry are absent. In a framework where the 
renormalization scale $\mu_{\chi}$ is proportional to $\chi$, the low 
energy effective action for the dilaton will be manifestly scale 
invariant. This symmetry allows derivative terms in the Lagrangian of 
the form
 \begin{equation}
\frac{1}{2} Z \partial_{\mu} \chi \partial^{\mu} \chi +
\frac{c}{\chi^4} \left(\partial_{\mu} \chi \partial^{\mu} \chi \right)^2
+ . \; . \; . 
 \end{equation}
For reasons that will become clear, we postpone rescaling $Z$ to one.
Crucially, however, in contrast to the effective theory of the NGB of 
spontaneously broken global symmetry, a non-derivative term in the 
potential is also allowed,
 \begin{equation}
V(\chi) = \frac{Z^2 \kappa_0}{4!} {\chi^4} \; .
 \end{equation}
The existence of this non-derivative term indicates that even in the 
absence of effects that explicitly violate conformal symmetry, there is 
a preferred value of $f = \langle \chi \rangle $. This is in sharp 
contrast to the case of a spontaneously broken global symmetry, where 
all points on the vacuum manifold are identical. In order to determine 
the location of the minimum, the effective potential must be obtained
and minimized.

In order to bring the theory into a standard form, we now go over to a 
scheme in which the renormalization scale $\mu$ is independent of 
$\chi$. In order to clarify the discussion, we first illustrate the 
procedure at one loop. We will obtain the Lagrangian for the low energy 
effective theory to this order, and use it to determine the effective 
potential and dilaton mass. We will then show how the result generalizes 
to arbitrary numbers of loops. It will be convenient to work in a 
mass-independent scheme, such as $\overline{{\rm MS}}$. We label $Z$ and 
the coupling constants $c$, $Z^2 \kappa_0$ etc. by $g_i$, where $i$ is 
an index. The $g_i$ are all dimensionless.

\subsubsection{One Loop Analysis}

At one loop, going over to a scheme where the renormalization scale 
$\mu$ is independent of $\chi$ is equivalent to evolving the parameters 
$g_i$ etc. from $\mu_{\chi}$ to $\mu$ using the renormalization group. 
Running the renormalization group leads to $g_i$ evolving into $g_i'$, 
where
 \begin{equation}
g_i' = g_i -
\frac{ {\rm d} g_i }{ {\rm d} \; {\rm log}\mu } {\rm log}
\left( \frac{\chi}{f} \right) \; .
 \end{equation}
To keep the analysis simple we focus on the case when all the $g_i$ are 
zero, except $Z$ and $Z^2 \kappa_0$. Then to this order, the potential 
for the dilaton takes the form
 \begin{equation}
V(\chi) = \left\{ Z^2 \kappa_0 -
\frac{ {\rm d} (Z^2 \kappa_0) }{ {\rm d} \; {\rm log}\mu } {\rm log}
\left( \frac{\chi}{f} \right) \right\}
\frac{\chi^4}{4!} \; .
\label{oneloopV}
 \end{equation}
Note that the potential is no longer manifestly scale invariant. In this 
theory at one loop order there is no wave function renormalization,
 \begin{equation}
\frac{ {\rm d} \; {\rm log} Z }{ {\rm d} \; {\rm log}\mu } = - 2 \gamma = 0 \; .
 \end{equation}
The derivative of $\kappa_0$ can be evaluated in perturbation theory, 
leading to
 \begin{equation}
\frac{ {\rm d} \kappa_0 }{ {\rm d} \; {\rm log}\mu } =
\frac{ 3 \kappa_0^2}{ 16 \pi^2} \; .
 \end{equation}
After using these expressions to replace the terms involving derivatives 
in the Lagrangian, we may choose to rescale $Z$ to one.

The conformal invariance of this Lagrangian can be made more transparent in 
a basis where all the mass scales are expressed as powers of the 
renormalization scale $\mu$, and all coupling constants are 
dimensionless. In such a basis, the dilaton kinetic term can be
written as
 \begin{equation}
\label{Zbarkineticterm}
\frac{1}{2} \bar{Z} \partial_{\mu} \chi \partial^{\mu} \chi \; ,
 \end{equation}
where $\bar{Z}$ is given to one loop order by
  \begin{equation}
\bar{Z} = Z - \frac{ {\rm d} Z }{ {\rm d} \; {\rm log}\mu }
{\rm log} \left( \frac{\mu}{f} \right) \; .
 \end{equation}
$\bar{Z}$, which is equal to $Z$ since wave function renormalization 
vanishes to this order, is a renormalization group invariant and does 
not change with $\mu$. At this point we choose to rescale $\bar{Z}$ to 
one.

In this basis the potential for the dilaton takes the form
 \begin{equation}
V(\chi) =
\left\{ \overline{\kappa}_0 -
\frac{ {\rm d} (Z^2 \kappa_0) }{ {\rm d} \; {\rm log}\mu } {\rm log}
\left( \frac{\chi}{\mu} \right) \right\}
\frac{\chi^4}{4!} \; .
 \end{equation}
where $\overline{\kappa}_0$ is given by
 \begin{equation}
\overline{\kappa}_0 = Z^2 {\kappa}_0 - 
\frac{ {\rm d} (Z^2 \kappa_0) }{ {\rm d} \; {\rm log}\mu }
{\rm log} \left( \frac{\mu}{f} \right) \; .
 \end{equation}
Note that $\overline{\kappa}_0$, like $\bar{Z}$, is independent of the 
renormalization scale $\mu$ to this loop order, as dictated by conformal 
invariance.

The next step is to determine the one loop effective potential. This can 
be computed from Eq.~(\ref{oneloopV}), after rescaling $Z$ to one, using 
the Coleman-Weinberg formula,
 \begin{equation}
\label{CWeffective}
V_{\rm eff} = V 
\pm \frac{1}{64 \pi^2} \sum_{i} {M_i^4} \left( {\rm log}
\frac{M_i^2}{\mu^2} - \frac{1}{2} \right) \; . 
 \end{equation}   
Here the sum is over the field dependent masses of all the states in the
theory, the sign being positive for bosons and negative for fermions.
This leads to 
 \begin{equation}
V_{\rm eff}(\chi_{cl}) = 
\left\{{\kappa}_0 -
\frac{3 \kappa_0^2}{32 \pi^2}
\left[{\rm log} \left( \frac{\mu^2}{\frac{1}{2}\kappa_0 f^2} \right) 
- \frac{1}{2} \right] \right\}
 \frac{\chi_{cl}^4}{4!}
 \end{equation}
The conformal invariance of the theory can be made clear by rewriting this in
terms of $\overline{\kappa}_0$. We obtain
 \begin{equation}
V_{\rm eff}(\chi_{cl}) = \frac{\hat{\kappa}_0}{4!} {\chi_{cl}^4} \; .
 \end{equation}
where $\hat{\kappa}_0$, given to one loop order by
 \begin{equation} 
\hat{\kappa}_0 = 
\overline{\kappa}_0 + \frac{3 \overline{\kappa}_0^2}{32 \pi^2} 
\left[ {\rm log} \left( \frac{\overline{\kappa}_0}{2} \right) - \frac{1}{2} 
\right] \; ,
 \end{equation}
is independent of the renormalization scale $\mu$, as 
required by conformal invariance.

Minimizing this effective potential, we find that the conformal symmetry 
breaking scale $\langle \chi \rangle = f$ is driven to zero, 
corresponding to unbroken conformal symmetry, if the sign of 
$\hat{\kappa}_0$ is positive. Alternatively, if $\hat{\kappa}_0$ is 
negative, $f$ is driven to infinite values, and conformal symmetry is 
never realized. Only if the value of $\hat{\kappa}_0$ is identically 
zero does the low energy effective theory possess a stable minimum, and 
a massless dilaton. In general setting $\hat{\kappa}_0 = 0$ is 
associated with tuning, since there is no symmetry reason to expect it 
to vanish.

\subsubsection{General Analysis} 

Although this result was obtained based on a one loop analysis, we now 
show that the same conclusion holds at arbitrary loop order. It can be 
verified that by replacing $g_i$ in the theory renormalized at $\mu_{\chi}$ 
by $g_i'$, where $g_i'$ is given by
 \begin{equation}
g_i' = g_i + 
\sum_{n = 1}^{\infty} \frac{(-1)^n}{n!}   
\frac{ {\rm d}^n g_i }{ {\rm d} \; {\rm log}\mu^n } \left[ {\rm log} 
\left( \frac{\chi}{f} \right) \right]^n \; ,
 \end{equation} 
we obtain a Lagrangian which is conformally invariant when renormalized 
at $\mu$. The higher terms in this series are to be determined 
self-consistently order by order in perturbation theory. The potential 
for the dilaton now takes the form
 \begin{equation}
V(\chi) = 
\left\{\sum_{n = 0}^{\infty} \frac{(-1)^n}{n!}
\frac{ {\rm d}^n (Z^2 \kappa_0) }{ {\rm d} \; {\rm log}\mu^n } \left[{\rm log}
\left( \frac{\chi}{f} \right) \right]^n \right\}
\frac{\chi^4}{4!} \; .
 \end{equation}
As expected, the Lagrangian does not possess a manifestly scale 
invariant form. We can choose to rescale $Z$ to one after the 
derivatives have been evaluated, but not before.

The conformal invariance of the theory can be made more transparent by 
going over to a basis where all the mass scales are expressed as powers 
of the renormalization scale $\mu$, and all coupling constants are 
dimensionless. In such a basis, $\bar{Z}$, the coefficient of the 
dilaton kinetic term, is given by
 \begin{equation}
\bar{Z} = \sum_{m = 0}^{\infty} \frac{(-1)^m}{m!} \frac{ {\rm d}^{m} Z}
{ {\rm d} \; {\rm log}\mu^{m} } \left[{\rm log}
\left(\frac{\mu}{f} \right) \right]^m \; .
 \end{equation}
$\bar{Z}$ does not change with the renormalization scale $\mu$, and we 
rescale it to one without loss of generality. The potential for the 
dilaton takes the form
  \begin{equation}
V(\chi) =  \left\{
\sum_{n = 0}^{\infty} \frac{(-1)^n}{n!}
\overline{\kappa}_{0,n} \left[{\rm log}
\left( \frac{\chi}{\mu} \right) \right]^n \right\}
\frac{\chi^4}{4!} \; ,
 \end{equation}
where $\overline{\kappa}_{0,n}$ is given by
 \begin{equation}
\overline{\kappa}_{0,n} 
= \sum_{m = 0}^{\infty} \frac{(-1)^m}{m!} \frac{ {\rm d}^{m+n} (Z^2 \kappa_0) 
}{ {\rm d} \; {\rm log}\mu^{m+n} } 
\left[{\rm log} \left(\frac{\mu}{f} \right) \right]^m \; .
 \end{equation}
The beta functions of all the $\overline{\kappa}_{0,n}$ vanish by 
construction. This is a reflection of the conformal invariance of this 
theory. Going forward, we denote the $\overline{\kappa}_{0,n}$ and all 
the other coupling constants in this basis by $\bar{g}_i$, where $i$ is 
an index. The beta functions of all the $\bar{g}_i$ vanish as a 
consequence of conformal invariance.

The next step is to obtain the effective potential $V_{\rm 
eff}(\chi_{cl})$ for this theory, and to minimize it. How is $V_{\rm 
eff}(\chi_{cl})$ to be determined? This time, rather than work directly 
from the Lagrangian, we employ the Callan-Symanzik equation for the 
effective potential,
 \begin{equation}
\label{CS0}
\left\{ \mu \frac{\partial}{\partial \mu} 
+ \beta_i \frac{\partial}{\partial \bar{g}_i} 
- \gamma \chi_{cl} \frac{\partial}{\partial \chi_{cl}} \right\}
V_{\rm eff}(\chi_{cl}, \bar{g}_i, \mu) = 0 \; .
 \end{equation} 
 For a conformal theory, the beta functions $\beta_i (\bar{g}_i)$ 
vanish. The anomalous dimension $\gamma$ of $\chi$, which represents the 
difference between its mass and scaling dimensions, is also zero. Then 
the Callan-Symanzik equation reduces to
 \begin{equation}
\mu \frac{\partial}{\partial \mu}
V_{\rm eff}(\chi_{cl}, \bar{g}_i, \mu) = 0 \; .
 \end{equation}
The effective potential is then constrained by dimensional analysis 
to be of the especially simple form
 \begin{equation}
V_{\rm eff} (\chi_{cl}) = \frac{\hat{\kappa}_0}{4!} {\chi_{cl}^4} \; ,
 \end{equation}
where $\hat{\kappa}_0$ is a constant that depends on the $\bar{g}_i$, 
but is independent of $\mu$. We see that the theory does not have a 
stable minimum unless $\hat{\kappa}_0 = 0 $, when the potential vanishes 
identically. The results of our one loop analysis are therefore 
confirmed.

\subsection{Incorporating Conformal Symmetry Violating Effects}

The situation changes if effects that explicitly break conformal 
symmetry are present in the theory. Consider an operator 
$\mathcal{O}(x)$ of scaling dimension $\Delta$ added to the Lagrangian,
 \begin{equation}
\mathcal{L} = \mathcal{L}_{\rm CFT} + \lambda_{\mathcal{O}} \mathcal{O}(x)
\; .
\label{deformation}
 \end{equation}
Under $ x \rightarrow x' = e^{-\omega} x$, the operator $\mathcal{O}(x) 
\rightarrow \mathcal{O}'(x') = e^{\omega \Delta} \mathcal{O} (x)$. It is 
convenient to define the dimensionless coupling constant 
$\hat{\lambda}_{\mathcal{O}} = \lambda_{\mathcal{O}} \mu^{\Delta - 4}$. 
We choose to normalize the operator $\mathcal{O}(x)$ such that 
$\hat{\lambda}_{\mathcal{O}}$ of order one corresponds to conformal 
symmetry violation becoming strong, so that it can no longer be treated 
as a perturbation on the conformal dynamics. This implies that if 
$\hat{\lambda}_{\mathcal{O}} \ll 1$, it satisfies the renormalization 
group equation
 \begin{equation}
\frac{ {\rm d} \; {\rm log} \hat{\lambda}_{\mathcal{O}}}
{{\rm d} \; {\rm log}\mu} = - (4 - \Delta) \; .
\label{RG}
 \end{equation}
We wish to determine the effect of this deformation on the form of the
low energy effective theory. In order to do this, note that for
small $\hat{\lambda}_{\mathcal{O}}$
the action remains formally invariant under $ x \rightarrow 
x' = e^{-\omega} x$ provided $\lambda_{\mathcal{O}}$ is taken to be a 
spurion that transforms as
 \begin{equation}
\label{spurion}
\lambda_{\mathcal{O}} \rightarrow \lambda'_{\mathcal{O}} = 
e^{\left(4 - \Delta \right) \omega} 
\lambda_{\mathcal{O}}.
 \end{equation}
This implies that the effective theory for $\chi$ will also respect 
conformal symmetry if $\lambda_{\mathcal{O}}$ is treated as a spurion 
that transforms in this way.

In determining the low energy effective theory for the dilaton it is 
again simplest to begin in a framework where the renormalization scale 
depends on the conformal compensator as $\mu_{\chi} = \mu \hat{\chi}$, 
since the Lagrangian is then manifestly scale invariant. The potential
for the dilaton is then
 \begin{equation}
V(\chi) = \frac{Z^2 \kappa_0}{4!} \chi^4 -
\sum_{n =1}^{\infty}
\frac{Z^{2 - n \epsilon/2} \kappa_n}{4!} {\lambda}_{\mathcal{O}}^n \;
\chi^{\left(4 - n \epsilon \right)} \; ,
\label{potn}
 \end{equation}
where $\epsilon$ is defined as $4 - \Delta$. The next step is to go over 
to a more conventional scheme where the renormalization scale $\mu$ is 
independent of $\chi$.

\subsubsection{One Loop Analysis}

In order to clarify the discussion we will first work in the limit that 
$\hat{\lambda}_{\mathcal{O}} \ll 1$ at scales $\mu$ of order $f$, and 
determine the vacuum structure and the dilaton mass to one loop order. 
We will then relax the assumption on $\hat{\lambda}_{\mathcal{O}}$ and 
also generalize the result to an arbitrary number of loops.

Keeping only the leading order term in $\hat{\lambda}_{\mathcal{O}}$, 
the potential for the dilaton Eq.~(\ref{potn}) simplifies to
 \begin{equation}
\label{pot1}
V(\chi) = \frac{Z^2 \kappa_0}{4!} \chi^4 
- \frac{Z^{\Delta/2} \kappa_1}{4!} \lambda_{\mathcal{O}} \: \chi^{\Delta} \; ,
 \end{equation}
where $\kappa_0$ and $\kappa_1$ are coupling constants. We can go over 
to a scheme where the renormalization scale is independent of $\chi$ by 
using the renormalization group. The potential then becomes, to one loop 
order,
 \begin{eqnarray}
V(\chi) &=&  \left\{ Z^2 \kappa_0 -
\frac{ {\rm d} (Z^2 \kappa_0) }{ {\rm d} \; {\rm log}\mu } {\rm log}
\left( \frac{\chi}{f} \right) \right\}  \frac{\chi^4}{4!} \\
&-& 
\left\{ Z^{\Delta/2} \kappa_1 -
\frac{ {\rm d} (Z^{\Delta/2} \kappa_1) }{ {\rm d} \; {\rm log}\mu } {\rm log}
\left( \frac{\chi}{f} \right) \right\} 
\frac{ \lambda_{\mathcal{O}} \: \chi^{\Delta}}{4!} \; . \nonumber
\label{oneloopVlambda}
 \end{eqnarray}
To keep the analysis simple we focus on the case when all the $g_i$ are 
zero, except $Z$, $Z^2 \kappa_0$ and $Z^{\Delta/2} \kappa_1$. This 
theory does not experience wave function renormalization at one loop, 
and therefore the derivatives of $Z$ in the expression above vanish. The 
derivatives of $\kappa_0$ and $\kappa_1$ can be evaluated in 
perturbation theory, leading to
 \begin{eqnarray}
\frac{ {\rm d} \kappa_0 }{ {\rm d} \; {\rm log}\mu } &=&
\frac{ 3 \kappa_0^2}{ 16 \pi^2} \nonumber \\
\frac{ {\rm d} \kappa_1 }{ {\rm d} \; {\rm log}\mu } &=&
\frac{ \Delta \left( \Delta - 1 \right) \kappa_1 \kappa_0}{ 32 \pi^2} 
\; .
 \end{eqnarray}

In order to understand how conformal symmetry is realized in this framework it
is useful to go over to a basis where all mass scales are expressed in 
terms of the renormalization scale $\mu$ and all coupling constants are
dimensionless. In this basis, $\bar{Z}$, the coefficient of the dilaton
kinetic term is given by
  \begin{equation}
\bar{Z} = Z - \frac{ {\rm d} Z }{ {\rm d} \; {\rm log}\mu }
{\rm log} \left( \frac{\mu}{f} \right) \; .
 \end{equation}
$\bar{Z}$ is a renormalization group invariant. The absence of wave 
function renormalization in this theory at one loop means that $Z = 
\bar{Z}$ to this order. We choose to rescale $\bar{Z}$ to one.

The potential for the dilaton takes the form
 \begin{eqnarray}
V(\chi) &=&  \left\{ \overline{\kappa}_0 -
\frac{ {\rm d} (Z^2 \kappa_0) }{ {\rm d} \; {\rm log}\mu } {\rm log}
\left( \frac{\chi}{\mu} \right) \right\} \frac{\chi^4}{4!} \nonumber \\
&-& 
\left\{ \overline{\kappa}_1 -
\frac{ {\rm d} (Z^{\Delta/2} \kappa_1) }{ {\rm d} \; {\rm log}\mu } {\rm log}
\left( \frac{\chi}{\mu} \right) \right\} 
\frac{\lambda_{\mathcal{O}} \chi^{\Delta}}{4!} \; .
 \end{eqnarray}
Here $\overline{\kappa}_0$ and $\overline{\kappa}_1$ are related to 
$Z$, ${\kappa}_0$ and ${\kappa}_1$ as
 \begin{eqnarray}
\overline{\kappa}_0 &=& Z^2 {\kappa}_0 -
\frac{ {\rm d} (Z^2 \kappa_0) }{ {\rm d} \; {\rm log}\mu }
{\rm log} \left( \frac{\mu}{f} \right) \nonumber \\
\overline{\kappa}_1 &=& {Z^{\Delta/2} \kappa}_1 -
\frac{ {\rm d} (Z^{\Delta/2} \kappa_1) }{ {\rm d} \; {\rm log}\mu }
{\rm log} \left( \frac{\mu}{f} \right) \; .
 \end{eqnarray}

Note that $\lambda_{\mathcal{O}}$ is being treated as a spurion, not as 
a coupling constant, and therefore continues to carry mass dimension $4 
- \Delta$, equal to its spurious scaling dimension. The beta functions 
of $\overline{\kappa}_0$ and $\overline{\kappa}_1$ can be seen to vanish 
to one loop order by construction. This is a consequence of the 
(spurious) conformal symmetry.

The next step is to obtain the effective potential for the theory at one 
loop order. We again use Eq.~(\ref{CWeffective}), after rescaling $Z$ to 
one, leading to
\begin{eqnarray}
\label{CWoneloop}
V_{\rm eff}(\chi_{cl}) &=&
\frac{\overline{\kappa}_0}{4!} \chi_{cl}^4 -
\frac{\overline{\kappa}_1}{4!} \lambda_{\mathcal{O}} 
\chi_{cl}^{\Delta} +
\\
&&  
\frac{\overline{\kappa}_0}{4!} 
\left[\frac{3 \overline{\kappa}_0}{32 \pi^2} \chi_{cl}^4
 - \frac{\lambda_{\mathcal{O}} \Delta \left( \Delta - 1 \right)
\overline{\kappa}_1} {64 \pi^2} \chi_{cl}^{\Delta} \right] 
\left[ \Sigma - \frac{1}{2} \right],
\nonumber 
\end{eqnarray}
where $\Sigma$ is defined as
 \begin{equation}
\Sigma = {\rm log} \left[\frac{\overline{\kappa}_0}{2} -
\frac{\overline{\kappa}_1}{4!} \Delta
\left( \Delta - 1 \right) \lambda_{\mathcal{O}} \chi_{cl}^{\Delta - 4}
\right]
\; .
 \end{equation}

The next step is to find the minimum of this potential, and to obtain 
the dilaton mass. For simplicity, we neglect the loop suppressed terms 
on the second line of Eq.~(\ref{CWoneloop}). We will later verify that 
including them does not alter our conclusions.  The tree level potential 
admits a minimum when
 \begin{equation}
f^{(\Delta - 4)} = \frac{4 \overline{\kappa}_0}{ \overline{\kappa}_1 
\lambda_{\mathcal{O}} \Delta} \; .
\label{naiveminimum} 
\end{equation}
The dilaton mass squared at the minimum, to this order, is given by
 \begin{equation}
m_\sigma^2 =  \frac{\overline{\kappa}_1}{4!} \lambda_{\mathcal{O}} 
\Delta (4 - \Delta) 
f^{\Delta - 2} = 4 \frac{\overline{\kappa}_0}{4!} (4 - \Delta )f^2
\; .
\label{naivedilatonmass}
 \end{equation}
If the conformal field theory is weakly coupled, the parameters 
$\overline{\kappa}_0, \overline{\kappa}_1 \ll (4 \pi)^2$, 
$\hat{\lambda}_{\mathcal{O}} \ll 1 \Rightarrow \lambda_{\mathcal{O}} 
f^{(\Delta - 4)} \ll 1$, and the effective theory of the dilaton we have 
obtained is valid. Corrections to Eqs.~(\ref{naiveminimum}) and 
(\ref{naivedilatonmass}) from the loop suppressed terms in 
Eq.~(\ref{CWoneloop}) can be seen to be small in this limit, and 
we are justified in neglecting them.

However, if the conformal field theory under consideration is strongly 
coupled, as in the theories of interest for electroweak symmetry 
breaking, the effective theory of the dilaton is also expected to be 
strongly coupled at the scale $\Lambda \sim 4 \pi f$. Then, in the 
absence of tuning, the parameters $\overline{\kappa}_0$ and 
$\overline{\kappa}_1$ are in general of order $(4 \pi)^2$ and, as is 
clear from Eq.~(\ref{naiveminimum}), the assumption that 
$\hat{\lambda}_{\mathcal{O}}$ is small at the scale $f$ is no longer 
self consistent. Furthermore, it follows from 
Eq.~(\ref{naivedilatonmass}) that the mass of the dilaton is of order 
the cutoff $\Lambda$ and so it is no longer a light state. The loop 
suppressed terms we have neglected in obtaining Eq.~(\ref{naiveminimum}) 
cannot alter this result. The conclusion to be drawn from this is that 
if a strongly coupled conformal field theory is explicitly broken by a 
relevant operator that becomes strong in the infrared, in general there 
is no reason to expect a light dilaton.

However, a closer study of Eq.~(\ref{naivedilatonmass}) reveals a very 
interesting feature. If the operator $\mathcal{O}$ is very close to 
marginal so that $(4 - \Delta) \ll 1$, then even for 
$\overline{\kappa}_0 \sim (4 \pi)^2$ the dilaton mass is parametrically 
smaller than the strong coupling scale $\Lambda$. It is straightforward 
to verify that this conclusion remains true even when the loop 
suppressed terms in Eq.~(\ref{CWoneloop}) are included in the analysis. 
This would suggest that in a scenario where the operator that breaks 
conformal symmetry is close to marginal, there is indeed a light dilaton 
in the effective theory. The dilaton mass depends on how close the 
dimension of $\mathcal{O}$ is to the exactly marginal value of 4, 
scaling as $m_{\sigma} \sim \sqrt{4 - \Delta}$.
 
This is potentially a very important result. In a large class of 
theories of interest for electroweak symmetry breaking, the operator 
that breaks conformal symmetry is close to marginal in order to ensure 
that there is a large hierarchy between the flavor scale (or Planck 
scale) and the electroweak scale. This result would imply that in all 
such theories the low energy spectrum includes a light dilaton! 
Unfortunately, the steps leading up to Eq.~(\ref{naivedilatonmass}) 
assumed that $\hat{\lambda}_{\mathcal{O}} \ll 1$. As is clear from 
Eq.~(\ref{naiveminimum}), this assumption is not valid in the strong 
coupling limit. In order to validate this conclusion, we must show that 
the result continues to hold when this assumption is relaxed, and is 
valid beyond one loop.

\subsubsection{General Analysis}

Extending the analysis beyond small $\lambda_{\mathcal{O}}$ involves 
incorporating two distinct effects. Firstly, if the coupling constant 
$\lambda_{\mathcal{O}}$ is not small, the scaling behavior of the 
operator ${\mathcal{O}}(x)$ is expected to receive corrections, and 
Eq.~(\ref{RG}) is in general no longer valid. Instead, the 
renormalization group equation takes on the more general form
 \begin{equation}
\label{generalRG}
\frac{ 
{\rm d} \; {\rm log} \hat{\lambda}_{\mathcal{O}}}{{\rm d} \; {\rm log}\mu} 
= - g(\hat{\lambda}_{\mathcal{O}}) \; ,
 \end{equation}
where $g(\hat{\lambda}_{\mathcal{O}})$ is in general a polynomial in 
$\hat{\lambda}_{\mathcal{O}}$,
 \begin{equation}
g(\hat{\lambda}_{\mathcal{O}}) 
= \sum_{n = 0}^{\infty} c_n \hat{\lambda}_{\mathcal{O}}^n \; ,
 \label{lambdaRGpolynomial}
 \end{equation}
that can be approximated by the lowest order term
 \begin{equation}
g(\hat{\lambda}_{\mathcal{O}}) = c_0 = (4 - \Delta) 
 \end{equation}
only in the limit when $\hat{\lambda}_{\mathcal{O}}$ is small. In 
general, in a strongly coupled conformal field theory, the coefficients 
$c_n$, $n \geq 1$ are expected to be of order one. (This is consistent 
with the expectation that all the terms in the series should become 
comparable when $\hat{\lambda}_{\mathcal{O}}$ is of order one.) This 
effect must be taken into account. Secondly, if 
$\hat{\lambda}_{\mathcal{O}}$ is not small, the higher order terms in 
Eq.~(\ref{potn}) are significant and must be included in our analysis.

While both these effects are important, the first has a particularly 
striking impact on the form of the low energy effective theory. The 
reason is that in this case, the leading order effect which is of order 
$(4 - \Delta)$ receives corrections that begin at order 
$\hat{\lambda}_{\mathcal{O}}$. Since in the theories of interest $(4 - 
\Delta) \ll 1$, these corrections can potentially become large even 
before $\hat{\lambda}_{\mathcal{O}}$ reaches its strong coupling value, 
significantly impacting the final result. This is most easily seen by 
going beyond the leading term in Eq.~(\ref{lambdaRGpolynomial}), while 
neglecting the higher order corrections in $\hat{\lambda}_{\mathcal{O}}$ 
that arise from other sources. Such an approximation is valid provided 
$\hat{\lambda}_{\mathcal{O}}$ at the breaking scale is significantly 
below its strong coupling value, $\hat{\lambda}_{\mathcal{O}} \ll 1$. We 
are interested in the region of parameter space where $ (4 - \Delta) < 
c_1 \hat{\lambda}_{\mathcal{O}}$, so that the corrections to the leading 
order term in Eq.~(\ref{lambdaRGpolynomial}) dominate. We postpone a 
more complete discussion that includes all the higher order effects in 
$\hat{\lambda}_{\mathcal{O}}$ till later in this section.
 
Integrating Eq~(\ref{generalRG}), it follows that 
$\mathcal{G}(\hat{\lambda}_{\mathcal{O}}) \mu^{-1}$ is a renormalization 
group invariant, where
 \begin{equation}
\label{Gdef}
 \mathcal{G}(\hat{\lambda}_{\mathcal{O}}) = {\rm exp}
\left( - \int \frac{d \hat{\lambda}_{\mathcal{O}}}
{\hat{\lambda}_{\mathcal{O}}}
 \frac{1}{g(\hat{\lambda}_{\mathcal{O}})}
\right) \; .
 \end{equation}
We can make the theory defined by Eq.~(\ref{deformation}) formally 
invariant under scale transformations by promoting 
$\hat{\lambda}_{\mathcal{O}}$ to a spurion that transforms as
 \begin{equation}
\hat{\lambda}_{\mathcal{O}}({\mu}) \rightarrow
\hat{\lambda}'_{\mathcal{O}}(\mu) = \hat{\lambda}_{\mathcal{O}}(\mu
e^{-\omega})
\label{lambdahatspurion}
 \end{equation}
under $x \rightarrow x' = e^{- \omega} x$. Under this transformation,
 \begin{equation}
\mathcal{G}(\hat{\lambda}_{\mathcal{O}}) \mu^{-1}
\rightarrow
\mathcal{G}(\hat{\lambda}'_{\mathcal{O}}) \mu^{-1}
=
e^{- \omega} \mathcal{G}(\hat{\lambda}_{\mathcal{O}}) \mu^{-1} \; .
\label{Gspurion}
 \end{equation}
The Lagrangian for the low energy effective theory must be invariant 
under this spurious scale transformation. Furthermore, it is restricted 
to terms involving positive integer powers of the spurion 
$\hat{\lambda}_{\mathcal{O}}$. Using Eq.~(\ref{Gdef}), it follows that 
the combination $\overline{\lambda}_{\mathcal{O}}$, defined as
 \begin{equation}
\overline{\lambda}_{\mathcal{O}}
= \hat{\lambda}_{\mathcal{O}}
\left[ 1 + g(\hat{\lambda}_{\mathcal{O}}) {\rm log} \mu  \right] \; ,
 \end{equation}
is invariant under infinitesimal changes in the renormalization scale $\mu$. 
It then follows from Eq.~(\ref{Gspurion}) that the object 
$\overline{\Omega}(\hat{\lambda}_{\mathcal{O}}, \chi/\mu)$, defined as
 \begin{equation}
\overline{\Omega}(\hat{\lambda}_{\mathcal{O}}, \chi/\mu) =  
\hat{\lambda}_{\mathcal{O}} 
\left[ 1 - g(\hat{\lambda}_{\mathcal{O}}) {\rm log} \left( \frac{\chi}{\mu}
\right) \right] \; ,
 \end{equation}
is invariant under infinitesimal (spurious) scale transformations. 
Furthermore, $\overline{\Omega}$ is a polynomial in 
$\hat{\lambda}_{\mathcal{O}}$. Lagrangians that are invariant under 
infinitesimal (spurious) scale transformations can be constructed using 
$\overline{\Omega}$.

For values of $\mu$ close to the symmetry breaking scale $f$ and 
$g(\hat{\lambda}_{\mathcal{O}}) \ll 1$, we can approximate 
$\overline{\Omega}$ as
 \begin{equation}
\overline{\Omega}(\hat{\lambda}_{\mathcal{O}}, \chi/\mu) =
\hat{\lambda}_{\mathcal{O}}
\left(\frac{\chi}{\mu} \right)^{- g(\hat{\lambda}_{\mathcal{O}})} \; .
 \end{equation}
To leading order in $\overline{\Omega}$ the potential for $\chi$ takes 
the form
 \begin{equation}
V(\chi) = \frac{\chi^4}{4!} \left( \kappa_0 - \kappa_1 
\overline{\Omega} \right) \; .
 \end{equation}
From this potential the dilaton mass at the minimum can be obtained as
 \begin{equation}
m_{\sigma}^2 = 4 \frac{\kappa_0}{4!} g(\hat{\lambda}_{\mathcal{O}}) f^2
 \end{equation} 
This expression for the dilaton mass is very similar to that in 
Eq.~(\ref{naivedilatonmass}), except in one important respect. We now 
see that it is the scaling behavior of the operator ${\mathcal{O}}$ at 
the breaking scale that determines the dilaton mass, rather than the 
scaling dimension of ${\mathcal{O}}$ in the far ultraviolet. In 
particular, this implies that for the dilaton of a spontaneously broken 
approximate conformal symmetry to be naturally light, it is not 
sufficient that $(4 - \Delta) \ll 1$, so that the operator that breaks 
the symmetry is close to marginal in the far ultraviolet. Instead, the 
requirement is that this operator be close to marginal at the symmetry 
breaking scale, so that $g(\hat{\lambda}_{\mathcal{O}}) \ll 1$ at $\mu = 
f$. Since in a general strongly coupled theory, 
$\hat{\lambda}_{\mathcal{O}}$, and therefore 
$g(\hat{\lambda}_{\mathcal{O}})$, are expected to be of order one at the 
breaking scale, this condition is not expected to be satisfied in the 
scenarios of interest for electroweak symmetry breaking (for which $4 - 
\Delta \ll 1$ suffices to address the flavor problem). This suggests 
that the existence of a light dilaton in these theories is associated 
with tuning (or more precisely, a coincidence problem). However, since 
the consistency of this analysis requires that 
$\hat{\lambda}_{\mathcal{O}} \ll 1$, it remains to show that including 
the higher order corrections in $\hat{\lambda}_{\mathcal{O}}$ that we 
have so far neglected, and which may be significant, does not 
affect this conclusion.

The next step is obtain the effective theory for the dilaton, 
consistently including all the higher order effects in 
$\hat{\lambda}_{\mathcal{O}}$. At this point it is convenient to 
separate out the corrections to the scaling behavior of ${\mathcal{O}}$
from these effects.
Recalling that 
$\epsilon$ is defined as $(4 - \Delta)$, we write
 \begin{equation}
g(\hat{\lambda}_{\mathcal{O}}) = \epsilon
+ \delta g(\hat{\lambda}_{\mathcal{O}})
\; ,
 \end{equation} 
 where $\delta g(\hat{\lambda}_{\mathcal{O}})$ represents the higher 
order corrections. In order to simplify our analysis we will consider 
the two cases $|\delta g(\hat{\lambda}_{\mathcal{O}})| < \epsilon $ and 
$|\delta g(\hat{\lambda}_{\mathcal{O}})| > \epsilon $, corresponding to 
the corrections to the scaling dimension of the operator ${\mathcal{O}}$ 
being smaller or larger than $\epsilon$ at the breaking scale, 
separately.

\bigskip
\noindent
{\underline{Limit When the Corrections Are Small}}
\\

We first consider the case when $ |\delta 
g(\hat{\lambda}_{\mathcal{O}})| < \epsilon $ at the breaking scale. In 
this limit we can simplify Eq.~(\ref{Gdef}) by performing a binomial 
expansion,
 \begin{equation}
\int d \hat{\lambda}_{\mathcal{O}}
\frac{1}{\hat{\lambda}_{\mathcal{O}} g(\hat{\lambda}_{\mathcal{O}})}
= \int \frac{d \hat{\lambda}_{\mathcal{O}}}
{\epsilon \hat{\lambda}_{\mathcal{O}}}
\left[ 1 - \frac{\delta g(\hat{\lambda}_{\mathcal{O}})}{\epsilon}
+\ldots \right] \; .
 \end{equation}
Then
 \begin{equation}
\mathcal{G}(\hat{\lambda}_{\mathcal{O}}) =
{\hat{\lambda}_{\mathcal{O}}}^{-1/{\epsilon}} \; 
{\rm exp} \left[ \int d \hat{\lambda}_{\mathcal{O}}
\frac{\delta g(\hat{\lambda}_{\mathcal{O}})}{\epsilon^2
\hat{\lambda}_{\mathcal{O}}}
+ \ldots \right] \; .
 \end{equation}
It follows from this that 
$\overline{\mathcal{G}}(\hat{\lambda}_{\mathcal{O}})$, defined as
 \begin{equation}
\overline{\mathcal{G}}(\hat{\lambda}_{\mathcal{O}})
= \left[\mathcal{G}(\hat{\lambda}_{\mathcal{O}})\right]^{- \epsilon}
\; ,
 \end{equation}
can be expanded as a polynomial in $\hat{\lambda}_{\mathcal{O}}$,
 \begin{equation}
\overline{\mathcal{G}}(\hat{\lambda}_{\mathcal{O}}) = 
\hat{\lambda}_{\mathcal{O}}  \left[ 1 - \int d \hat{\lambda}_{\mathcal{O}}
\frac{\delta g(\hat{\lambda}_{\mathcal{O}})}
{\epsilon \hat{\lambda}_{\mathcal{O}}}
+ \ldots \right] \; .
 \end{equation}
Then the object $\overline{\mathcal{G}}(\hat{\lambda}_{\mathcal{O}}) 
\mu^{\epsilon}$, which we denote by $\overline{\lambda}_{\mathcal{O}}$, 
is a renormalization group invariant that can be expanded as a 
polynomial in $\hat{\lambda}_{\mathcal{O}}$. It follows from 
Eq.~(\ref{Gspurion}) that the theory above the breaking scale is 
formally invariant under scale transformations, $ x \rightarrow x' = 
e^{-\omega} x$, provided $\overline{\lambda}_{\mathcal{O}}$ is taken to 
be a spurion that transforms as
 \begin{equation}
\overline{\lambda}_{\mathcal{O}} \rightarrow
\overline{\lambda}'_{\mathcal{O}} =
e^{\epsilon \omega}
\overline{\lambda}_{\mathcal{O}}.
 \end{equation}
The effective theory for $\chi$ will then respect conformal symmetry if
$\overline{\lambda}_{\mathcal{O}}$ is treated as a spurion that
transforms in this way. Note that this spurious transformation is
identical to that of ${\lambda}_{\mathcal{O}}$, Eq.~(\ref{spurion}), in
the case of small $\hat{\lambda}_{\mathcal{O}}$. 

Consider the object $\Omega(\overline{\lambda}_{\mathcal{O}}, \chi)$,
defined as
 \begin{equation}
\Omega(\overline{\lambda}_{\mathcal{O}}, \chi) = 
\overline{\lambda}_{\mathcal{O}} \chi^{- \epsilon} \; .
 \end{equation}
By construction, $\Omega$ is invariant under (spurious) scale 
transformations. Furthermore, in the regime $ |\delta 
g(\hat{\lambda}_{\mathcal{O}})| < \epsilon $, it can be expanded as a 
polynomial in $\hat{\lambda}_{\mathcal{O}}$. $\Omega$ is useful in 
constructing the general Lagrangian for the low energy theory.

In a framework where the renormalization scale depends on the conformal 
compensator as $\mu_{\chi} = \mu \hat{\chi}$, the potential for $\chi$ 
takes the form 
 \begin{equation}
V(\chi) = \frac{Z^2 \chi^4}{4!} \left[
\kappa_0 - \sum_{n = 1}^{\infty}  \kappa_n
\Omega^n(\overline{\lambda}_{\mathcal{O}}, \sqrt{Z} \chi) 
 \right] \; .
\label{potomega}
 \end{equation}
This simplifies to the form of Eq~(\ref{potn}), but with 
${\lambda}_{\mathcal{O}}$ replaced by $\overline{\lambda}_{\mathcal{O}}$,
 \begin{equation}
V(\chi) = \frac{Z^2 \kappa_0}{4!} \chi^4 -
\sum_{n =1}^{\infty}
\frac{Z^{2 - n \epsilon/2} \kappa_n}{4!} \overline{\lambda}_{\mathcal{O}}^n \;
\chi^{\left(4 - n \epsilon \right)} \; .
 \end{equation}
Going over to a more conventional scheme where the renormalization scale 
$\mu$ is independent of $\chi$, $V(\chi)$ becomes
 \begin{eqnarray}
\label{chipot1.2}
&&
\sum_{m = 0}^{\infty} \frac{(-1)^m}{m!}
\frac{ {\rm d}^m (Z^2 \kappa_0)}{ {\rm d} \; {\rm log} \mu^m }
\left[{\rm log} \left( \frac{\chi}{f} \right) \right]^m
\frac{\chi^4}{4!} - \\
&&
\sum_{n = 1}^{\infty}  \sum_{m = 0}^{\infty} \frac{(-1)^m}{m!}
\frac{ {\rm d}^m (Z^{2 - n \epsilon/2} \kappa_n)}{ {\rm d} \; {\rm log} \mu^m }
\left[{\rm log} \left( \frac{\chi}{f} \right) \right]^m
\frac{\overline{\lambda}_{\mathcal{O}}^n \chi^{4 - n \epsilon}}{4!} 
\nonumber\:.
 \end{eqnarray}
We can choose to rescale $Z$ to one, but only after the derivatives 
above have been evaluated.

The (spurious) conformal symmetry of the theory can be made more 
transparent in a basis where all mass scales are expressed in terms of 
the renormalization scale $\mu$ and all coupling constants are 
dimensionless. In this basis, $\bar{Z}$, the coefficient of the
dilaton kinetic term, is given by 
 \begin{equation}
\bar{Z}
= \sum_{m = 0}^{\infty} \frac{(-1)^m}{m!} \frac{ {\rm d}^{m} Z}
{ {\rm d} \; {\rm log}\mu^{m} } \left[{\rm log}
\left(\frac{\mu}{f} \right) \right]^m \; .
 \end{equation}
$\bar{Z}$ is independent of the renormalization scale $\mu$. We again 
choose to set it to one. The potential for the dilaton now takes the 
form
  \begin{eqnarray}
V(\chi) &=& 
\sum_{m = 0}^{\infty} \frac{(-1)^m}{m!}
{\overline{\kappa}_{0,m}} \left[{\rm log}
\left( \frac{\chi}{\mu} \right) \right]^m \frac{\chi^4}{4!} \\
&-&
\sum_{n = 1}^{\infty}
\sum_{m = 0}^{\infty} \frac{(-1)^m}{m!}
{\overline{\kappa}_{n,m}} \left[{\rm log}
\left( \frac{\chi}{\mu} \right) \right]^m 
\frac{\overline{\lambda}_{\mathcal{O}}^n \chi^{4 - n \epsilon}}{4!}
 \nonumber \; 
 \end{eqnarray}
where the couplings constants $\overline{\kappa}_{n,m}$ are given by
  \begin{equation}
\sum_{r = 0}^{\infty} \frac{(-1)^r}{r!} \frac{{\rm d}^{m+r} 
(Z^{2 - n \epsilon/2} \kappa_n)
}{ {\rm d} \; {\rm log}\mu^{m+r} } \left[{\rm log}
\left(\frac{\mu}{f} \right) \right]^r \; .
 \end{equation}
The beta functions of all the $\overline{\kappa}_{n,m}$ vanish by 
construction, reflecting the (spurious) conformal invariance of the 
theory.

The final step is to determine the form of the effective potential.
We will again use the Callan-Symanzik equation for the effective
potential,
 \begin{equation}
\left\{ \mu \frac{\partial}{\partial \mu}
+ \beta_i \frac{\partial}{\partial \bar{g}_i}
- \gamma_{\phi_{\alpha}} \phi_{\alpha} \frac{\partial}
{\partial \phi_\alpha} \right\}
V_{\rm eff}(\phi_{\alpha}, \bar{g}_i, \mu) = 0 \; .
\label{CSgen}
 \end{equation} 
Here the index $\alpha$ runs over the fields in the theory, namely 
$\chi_{cl}$ and $\hat{\lambda}_{\mathcal{O}}$. The beta functions 
$\beta_i (\bar{g}_i)$ vanish as a consequence of the (spurious) conformal 
symmetry, as does the anomalous dimension of $\chi$. The anomalous 
dimension of $\hat{\lambda}_{\mathcal{O}}$ is 
$g(\hat{\lambda}_{\mathcal{O}})$, the difference between its scaling 
dimension and mass dimension. Then the Callan-Symanzik equation reduces 
to
 \begin{equation}
\left\{ 
\mu \frac{\partial}{\partial \mu}
- g(\hat{\lambda}_{\mathcal{O}}) \hat{\lambda}_{\mathcal{O}} 
\frac{\partial}{\partial \hat{\lambda}_{\mathcal{O}}}
\right\}
V_{\rm eff}(\chi_{cl}, \hat{\lambda}_{\mathcal{O}}, \bar{g}_i, \mu) 
= 0 \; .
 \end{equation}
Making a change of variable from $\hat{\lambda}_{\mathcal{O}}$ to
$\overline{\lambda}_{\mathcal{O}}$, this becomes simply
 \begin{equation}
\mu \frac{\partial}{\partial \mu}
V_{\rm eff}(\chi_{cl}, \overline{\lambda}_{\mathcal{O}}, \bar{g}_i, \mu)
= 0 \; .
 \end{equation}
Dimensional analysis constrains the solution to be of the form
 \begin{equation}
V_{\rm eff}(\chi_{cl}) = \frac{1}{4!} \chi_{cl}^4
\left[ {\hat{\kappa}_0}
- \mathcal{F}(\overline{\lambda}_{\mathcal{O}}
\chi_{cl}^{- \epsilon}) \right] \; ,
 \end{equation} 
where $\hat{\kappa}_0$ is a constant that depends on the couplings 
$\bar{g}_i$ but not on $\overline{\lambda}_{\mathcal{O}}$. The form of 
the function $\mathcal{F}(\Omega)$ cannot be determined from symmetry 
considerations alone, but depends on the dynamics of the conformal field 
theory under consideration, and on the operator $\mathcal{O}$. For 
values of $\Omega$ less than one by a factor of at least a few, 
corresponding to $\hat{\lambda}_{\mathcal{O}}$ being below its strong 
coupling value at the symmetry breaking scale $f$, $\mathcal{F}(\Omega)$ 
can be computed in perturbation theory. In general it is not a 
polynomial in $\Omega$, as can be seen from Eq.~(\ref{CWoneloop}).

Minimizing the effective potential we find the condition that determines the
symmetry breaking scale $f$,
 \begin{equation}
4 \hat{\kappa}_0 - 4 \mathcal{F}(\overline{\lambda}_{\mathcal{O}}
f^{- \epsilon}) + \epsilon \overline{\lambda}_{\mathcal{O}}
f^{- \epsilon} \mathcal{F}'(\overline{\lambda}_{\mathcal{O}}
f^{- \epsilon}) = 0 \; .
 \label{minimum}
 \end{equation}
The dilaton mass squared depends on the second derivative of the 
effective potential at the minimum, which is given by
 \begin{equation}
\frac{\partial^2 V_{\rm eff}}{\partial \sigma^2} = 
\frac{1}{4!} f^2 
\left\{ 4 \epsilon  
\overline{\lambda}_{\mathcal{O}} f^{- \epsilon}
\mathcal{F}'(\overline{\lambda}_{\mathcal{O}} f^{- \epsilon})
\right\} \; .
\label{dilatonmass}
 \end{equation}
Here we are neglecting effects of order $\epsilon^2$. We see from this 
that even at strong coupling, corresponding to $\hat{\kappa}_0 \sim (4 
\pi)^2$, if the function $\mathcal{F}(\Omega)$ satisfies the condition 
$\mathcal{F}(\Omega) \gtrsim \Omega \mathcal{F}'(\Omega)$ at the 
minimum, the dilaton mass is suppressed by a factor of $\sqrt{\epsilon}$ 
relative to the strong coupling scale $4 \pi f$, and therefore remains 
light. The question is whether this condition on the function 
$\mathcal{F}(\Omega)$ is indeed satisfied in a general strongly coupled 
conformal field theory, for an arbitrary marginal operator 
$\mathcal{O}$. Unfortunately, in the absence of additional information 
about the function $\mathcal{F}(\Omega)$, we cannot establish such a 
conclusion. At the minimum, the value of $\Omega$ is equal to that of 
$\hat{\lambda}_{\mathcal{O}}$ evaluated at the symmetry breaking scale 
$f$. For $\Omega$ of order one, corresponding to 
$\hat{\lambda}_{\mathcal{O}}$ close to its strong coupling value, we 
expect that $\mathcal{F}(\Omega)$ is of order $(4 \pi)^2$, but its 
functional form is completely unknown.

However, there exists a class of strongly coupled theories where the 
condition $\mathcal{F}(\Omega) \sim \Omega \mathcal{F}'(\Omega)$ is 
satisfied, and the dilaton is light. In the region of parameter space 
where $\hat{\lambda}_{\mathcal{O}}$ and $\hat{\kappa}_0$ are below their 
strong coupling values, the form of the function $\mathcal{F}(\Omega)$ 
can be determined from perturbation theory. In this regime it is 
dominated by the term linear in $\Omega$ in Eq.~(\ref{potomega}), since 
the other terms are loop suppressed or higher order in 
$\hat{\lambda}_{\mathcal{O}}$. Now, we expect that there exist strongly 
coupled conformal field theories where the parameter $\hat{\kappa}_0$ is 
below its natural strong coupling value by a factor of order a few. This 
is quite natural, requiring at most mild tuning. From the minimization 
condition it follows that in such theories, symmetry breaking is 
realized for values of $\mathcal{F}(\Omega)$ that correspond to values 
of $\Omega$, and therefore $\hat{\lambda}_{\mathcal{O}}$, that lie below 
their strong coupling values by roughly the same factor. Since 
$\mathcal{F}(\Omega)$ is linear in $\Omega$ in this regime, the 
condition $\mathcal{F}(\Omega) \sim \Omega \mathcal{F}'(\Omega)$ is 
satisfied at the minimum. Therefore in this class of theories the 
conclusion $m_{\sigma} \sim \sqrt{\epsilon}$ is valid, and the dilaton 
is light.

Since this analysis is restricted to the region of parameter space where 
$ |\delta g(\hat{\lambda}_{\mathcal{O}})| < \epsilon $, it is important 
to understand the circumstances under which this condition is satisfied. 
One possibility is that $\mathcal{O}$ is a protected operator, so that 
all the coefficients $c_n$ in the polynomial expansion of 
$g(\hat{\lambda}_{\mathcal{O}})$ are of order $\epsilon$. The operator 
$\mathcal{O}$ is then close to marginal for any value of 
$\hat{\lambda}_{\mathcal{O}}$. An example is a theory where the 
parameter $\hat{\lambda}_{\mathcal{O}}$ corresponds to a fixed line, 
while the parameter $\epsilon$ is associated with the coefficient of an 
operator that is very close to marginal (for all 
$\hat{\lambda}_{\mathcal{O}}$) and which lifts the fixed line. However, 
theories that admit such protected operators are clearly rather special. 
There is no reason to expect the condition $c_n \lesssim \epsilon$ to be 
satisfied by an arbitrary marginal operator $\mathcal{O}$ in a general 
conformal field theory.

Another possibility is that the parameter $\hat{\kappa}_0$ lies 
significantly below its natural strong coupling value so that symmetry 
breaking is realized for values of $\hat{\lambda}_{\mathcal{O}}$ less 
than $\epsilon$. The condition $|\delta g(\hat{\lambda}_{\mathcal{O}})| 
< \epsilon $ can then be satisfied. Since in this regime 
$\mathcal{F}(\Omega)$ is dominated by the term linear in $\Omega$, 
$\mathcal{F}(\Omega) \sim (4 \pi)^2 \Omega \sim (4 \pi)^2 
\hat{\lambda}_{\mathcal{O}}$, it follows from Eq.~(\ref{minimum}) that 
the condition $\hat{\lambda}_{\mathcal{O}} < \epsilon$ translates into 
$\hat{\kappa}_0/(4 \pi)^2 \lesssim \epsilon$. It follows from 
Eq.~(\ref{dilatonmass}) and the minimization condition 
Eq.~(\ref{minimum}) that in this regime the dilaton mass scales as 
$\sqrt{\hat{\kappa}_0 \epsilon}$, and therefore receives additional 
suppression from the fact that $\hat{\kappa}_0$ is small. However, small 
values of $\hat{\kappa}_0$ are associated with tuning, and so this 
condition is not expected to be satisfied in a general conformal field 
theory. However, in the case of small hierarchies, such as between the 
flavor scale and the weak scale, values of $\epsilon$ as large as 1/5 
can still serve to address the problem. It follows from this that in 
such a theory, a dilaton mass a factor of 5 below the strong coupling 
scale can be realized for $\hat{\kappa}_0$ a factor of 5 below its 
natural strong coupling value. Since the tuning scales with 
$\hat{\kappa}_0$, this theory need only be tuned at the level of 1 part 
in 5 (20\%). This is to be contrasted with the case of a (non-pNGB) 
composite scalar of the same mass, which is tuned at the level of 1 part 
in 25 (4\%). We see that although this scenario is tuned, the tuning is 
mild, scaling with the mass of the dilaton rather than the square of its 
mass.

\bigskip
\noindent
{\underline{Limit When Corrections Are Large}}
\\

We now turn our attention to the case when the corrections to the 
scaling behavior of $\mathcal{O}$ are large in the neighborhood of the 
breaking scale, so that $|\delta g(\hat{\lambda}_{\mathcal{O}})| > 
\epsilon $. For simplicity, we will work in the limit that the 
renormalization group evolution of ${\rm log}\; 
\hat{\lambda}_{\mathcal{O}}$ close to the breaking scale is dominated by 
the term linear in $\hat{\lambda}_{\mathcal{O}}$ so that
 \begin{equation}
\frac{
{\rm d} \; {\rm log} \hat{\lambda}_{\mathcal{O}}}{{\rm d} \; {\rm log}\mu}
= - c_1 \hat{\lambda}_{\mathcal{O}} \; ,
 \end{equation}
Integrating this equation we find that 
$\mathcal{G}(\hat{\lambda}_{\mathcal{O}})$ is now given by
 \begin{equation}
 \mathcal{G}(\hat{\lambda}_{\mathcal{O}}) = {\rm exp}
\left( \frac{1}{c_1 \hat{\lambda}_{\mathcal{O}}}
\right) \; .
 \end{equation}
Since $\mathcal{G}(\hat{\lambda}_{\mathcal{O}}) \mu^{-1}$ is a 
renormalization group invariant, it follows that 
$\overline{\lambda}_{\mathcal{O}}$, now defined as
 \begin{equation}
\overline{\lambda}_{\mathcal{O}} =
\frac{\hat{\lambda}_{\mathcal{O}}}
{1 - c_1 \hat{\lambda}_{\mathcal{O}} \; {\rm log} \mu}
 \end{equation}
is also a renormalization group invariant. Once 
$\hat{\lambda}_{\mathcal{O}}$ is promoted to a spurion as in 
Eq~(\ref{lambdahatspurion}), the object
 \begin{equation}
\Omega(\overline{\lambda}_{\mathcal{O}}, \chi) = 
\frac{\overline{\lambda}_{\mathcal{O}}}
{1 + c_1 \overline{\lambda}_{\mathcal{O}} \; {\rm log} \chi}
 \end{equation}
is invariant under (spurious) scale transformations. Furthermore, at 
scales $\mu$ close to $\langle \chi \rangle = f$, it can be expanded as 
a polynomial in $\hat{\lambda}_{\mathcal{O}}$. The Lagrangian for the 
low energy effective theory can be constructed using $\Omega$.

In a scheme where the renormalization scale is proportional to $\chi$,
$\mu_{\chi} = \mu \hat{\chi}$, the potential takes the form
 \begin{equation}
V(\chi) = \frac{Z^2 \chi^4}{4!} \left[ 
\kappa_0 - \sum_{n = 1}^{\infty}  \kappa_n
\Omega^n(\overline{\lambda}_{\mathcal{O}}, \sqrt{Z} \chi)
\right] \; .
 \end{equation}
It is straightforward to go over to a scheme where the renormalization 
scale $\mu$ is independent of $\chi$, and where all mass parameters in 
the Lagrangian are expressed as powers of $\mu$. As before, the 
dimensionless coupling constants $\bar{g}_i$ in such a scheme are 
independent of the renormalization scale $\mu$. 

The effective potential for the low energy effective theory can once 
again be determined from the Callan-Symanzik equation, Eq~(\ref{CSgen}). 
The anomalous dimension of $\chi$ vanishes while that of the spurion 
$\hat{\lambda}_{\mathcal{O}}$ is given by $ 
g(\hat{\lambda}_{\mathcal{O}}) = c_1 \hat{\lambda}_{\mathcal{O}}$. As a 
consequence the Callan-Symanzik equation reduces to
 \begin{equation}
\left\{
\mu \frac{\partial}{\partial \mu}
- c_1 \hat{\lambda}_{\mathcal{O}}^2
\frac{\partial}{\partial \hat{\lambda}_{\mathcal{O}}}
\right\}
V_{\rm eff}(\chi_{cl}, \hat{\lambda}_{\mathcal{O}}, \bar{g}_i, \mu)
= 0 \; .
 \end{equation}
Making the change of variable from $\hat{\lambda}_{\mathcal{O}}$ to
$\overline{\lambda}_{\mathcal{O}}$, this simplifies to
 \begin{equation}
\mu \frac{\partial}{\partial \mu}
V_{\rm eff}(\chi_{cl}, \overline{\lambda}_{\mathcal{O}}, \bar{g}_i, \mu)
= 0 \; .
 \end{equation}
Dimensional analysis constrains the solution to be of the form
 \begin{equation}
V_{\rm eff}(\chi_{cl}) = \frac{1}{4!} \chi_{cl}^4
\left\{ \hat{\kappa}_0
- \mathcal{F}[\Omega(\overline{\lambda}_{\mathcal{O}},
\chi_{cl})] \right\} \; .
 \end{equation}
The form of the function $\mathcal{F}[\Omega]$ cannot be determined from 
symmetry considerations alone, but depends on the dynamics of the 
conformal field theory under consideration, and on the operator 
$\mathcal{O}$. For values of $\Omega$ less than one by at least a factor 
of a few, corresponding to $\hat{\lambda}_{\mathcal{O}}$ being below its 
strong coupling value at the symmetry breaking scale, 
$\mathcal{F}(\Omega)$ can be computed in perturbation theory.

Minimizing the effective potential we obtain the condition that determines 
$f$,
 \begin{equation}
4 \hat{\kappa}_0 - 4 \mathcal{F}[\Omega]
+ c_1 \Omega^2 
\mathcal{F}'[\Omega] = 0 \; .
\label{lastminimum}
 \end{equation}
The dilaton mass squared depends on the second derivative of the
effective potential at the minimum, which can be determined as
 \begin{equation}
\frac{\partial^2 V_{\rm eff}}{\partial \sigma^2} =
\frac{c_1}{4!} f^2 \Omega^2
\left[  
\left( 4 - 2 c_1 \Omega \right) \mathcal{F}'
- c_1 \Omega^2 \mathcal{F}''
\right] \; .
\label{lastmass}
 \end{equation}
 
Once again we focus on theories where the parameter $\hat{\kappa}_0$ is 
below its natural strong coupling value by some factor, which could be 
as small as a few. In such theories, symmetry breaking is realized for 
values of $\hat{\lambda}_{\mathcal{O}}$ that are below its strong 
coupling value. In this limit the form of the effective potential can be 
determined in perturbation theory, and $\mathcal{F}(\Omega)$ is 
dominated by the term linear in $\Omega$. Then at the minimum the 
condition $\mathcal{F} [\Omega] \sim \Omega \mathcal{F}' [\Omega]$ is 
satisfied. Noting that at the minimum the value of $\Omega$ is equal to 
that of $\hat{\lambda}_{\mathcal{O}}$ at the scale $f$, it follows from 
Eq.~(\ref{lastminimum}) and Eq.~(\ref{lastmass}) that the dilaton mass 
squared scales as $m_{\sigma}^2 \sim c_1 \hat{\lambda}_{\mathcal{O}} 
\hat{\kappa}_0$.

We see from this that the dilaton mass depends on the scaling behavior 
of the operator ${\mathcal{O}}$ at the symmetry breaking scale, and 
therefore on the value of $c_1 \hat{\lambda}_{\mathcal{O}}$. Since the 
minimization condition Eq.~(\ref{lastminimum}) relates $\Omega$ (and 
therefore $\hat{\lambda}_{\mathcal{O}}$ at the breaking scale) to 
$\hat{\kappa}_0$, for $c_1$ of order its natural value of one we have 
that the dilaton mass squared scales as $\hat{\kappa}_0^2$. This 
suggests that for $\hat{\kappa}_0$ of order its natural strong coupling 
value the dilaton mass lies near the cutoff of the theory, and it is not 
a light state. It follows from this that in general, the spectrum of a 
conformal field theory broken by an arbitrary marginal operator that 
grows strong in the infrared does not include a light dilaton.

The low energy effective theory will however contain a light dilaton if 
the parameter $\hat{\kappa}_0$ lies significantly below its natural 
strong coupling value. In general, this involves tuning, since this 
condition is not expected to be satisfied in an arbitrary conformal 
field theory. However, the tuning is mild, scaling with $\hat{\kappa}_0$ 
and therefore as the mass of the dilaton, so that a dilaton that lies a 
factor of 5 below the strong coupling scale is only tuned at the level 
of 1 part in 5 (20 \%).
 
It follows from this discussion that in strongly coupled theories where 
an approximate conformal symmetry is spontaneously broken, the low 
energy spectrum includes a light dilaton if the operator that breaks the 
symmetry is close to marginal at the breaking scale. This condition is 
in general not expected to be satisfied by the theories of interest for 
electroweak symmetry breaking, and so the presence of a light dilaton in 
these theories is associated with tuning. However, the tuning is mild, 
scaling as the mass of the dilaton rather than as the square of its 
mass.

\section{Dilaton Interactions in a Conformal SM}

In the limit that conformal invariance is exact, the form of the dilaton 
interactions with SM fields in the low energy effective theory is fixed 
by the requirement that the symmetry be realized nonlinearly. However, 
in the scenario of interest, we expect significant deviations from exact 
conformal invariance because effects associated with the operator 
${\mathcal{O}}$ that violate the symmetry are large at the breaking 
scale. It is crucial to understand the size of these effects, and the 
extent to which the predictions of the theory with exact conformal 
invariance are affected.

In this section, we consider a scenario where the SM gauge bosons and 
matter fields are all composites of a strongly interacting conformal 
sector that breaks electroweak symmetry dynamically, and there is no 
light Higgs. The AdS/CFT correspondence relates this scenario to 
Higgsless Randall-Sundrum models where the SM matter and gauge fields 
are localized on the infrared brane. The couplings of the dilaton to the 
SM fields in such a framework have been 
determined~\cite{arXiv:0708.1463}, and agree with earlier results for 
the couplings of the radion to brane-localized fields in Randall-Sundrum 
models~\cite{Goldberger:1999uk, Csaki:1999mp, Giudice:2000av, 
Csaki:2000zn}. Several authors have studied the question of 
distinguishing the dilaton from the Higgs at the LHC in such a 
scenario~\cite{arXiv:0803.2040, Barger:2011nu, Coleppa:2011zx}, see 
also~\cite{Campbell:2011iw}. We will study the corrections to the 
dilaton couplings in this scenario when effects associated with the 
operator ${\mathcal{O}}$ that explicitly violates conformal symmetry are 
incorporated.

We begin by considering the dilaton couplings to the $W$ and $Z$ 
gauge bosons. We choose to work in a basis where we write all gauge 
kinetic terms in the form
 \begin{equation}
-\frac{1}{4 g^2} F_{\mu \nu} F^{\mu \nu} \; .
 \end{equation}
In the absence of conformal symmetry violating effects, the couplings of 
the dilaton to the $W$ are such as to compensate for the breaking of 
conformal invariance by the gauge boson mass term. In unitary gauge 
these take the form
 \begin{equation}
\left( \frac{\chi}{f} \right)^2
\frac{m_W^2}{g^2} W_{\mu}^+ W^{\mu -}
\label{dilatontoW}
 \end{equation}
in the Lagrangian. Here $m_W$ is the $W$ gauge boson mass. Expanding the 
compensator $\chi = f e^{\sigma/f}$ out in terms of $\sigma$ to leading 
order in inverse powers of $f$, we find for the dilaton couplings
 \begin{equation}
2 \frac{\sigma}{f} \frac{m_W^2}{g^2} W_{\mu}^+ W^{\mu -} \; .
\label{basic}
 \end{equation}

Next we consider the corrections to the dilaton couplings when conformal 
symmetry violating effects are included. We will focus on the case when 
the corrections to the scaling behavior of $\mathcal{O}$ are small, so 
that $|\delta g(\hat{\lambda}_{\mathcal{O}})|$ is less than $\epsilon$ 
at the breaking scale. We will later argue that the same conclusions are 
obtained in the limit when $|\delta g(\hat{\lambda}_{\mathcal{O}})|$ is 
greater than $\epsilon$.

The presence of conformal symmetry violating effects allows additional 
two derivative terms in the dilaton action,
 \begin{equation}
\frac{1}{2} \left[ 1 + \sum_{n =1}^{\infty}
\alpha_{\chi, n} \overline{\lambda}_{\mathcal{O}}^n \;
\chi^{\left( - n \epsilon \right)} \right] 
\partial_{\mu} \chi \partial^{\mu} \chi \; .
\label{chikinlambda}
 \end{equation}
The dimensionless parameters $\alpha_{\chi, n}$ depend both on the 
operator $\mathcal{O}$ and the specific conformal field theory under 
consideration. They are expected to be of order one. These new terms 
contribute to the dilaton kinetic term, which now becomes
 \begin{equation}
\frac{1}{2} \left[ 1 + \sum_{n =1}^{\infty}
\alpha_{\chi, n} \overline{\lambda}_{\mathcal{O}}^n \;
f^{\left( - n \epsilon \right)} \right]
\partial_{\mu} \sigma \partial^{\mu} \sigma \; .
 \end{equation}
When $\sigma$ is rescaled to make the dilaton kinetic term canonical, we 
see that the effective impact of these terms is to alter the effective 
value of $f$ in Eq.~(\ref{basic}), while leaving the form of the 
interaction unchanged. More generally, it follows that corrections to 
the dilaton kinetic term from conformal symmetry violating effects do 
not alter the form of the dilaton couplings to the SM fields. Instead, 
to leading order in $\sigma/f$, they lead to a universal rescaling in 
the effective value of $f$, leaving the relative strengths of the 
dilaton couplings to the various SM fields unchanged. Since to the order 
we are working this effect can be entirely absorbed into the parameter 
$f$, we will not consider it further.

The gauge kinetic term also receives corrections from conformal symmetry 
violating effects. It now takes the form
 \begin{equation}
-\frac{1}{4 \hat{g}^2}  \left[ 1 + \sum_{n =1}^{\infty}
\alpha_{W, n} \overline{\lambda}_{\mathcal{O}}^n \;
\chi^{\left( - n \epsilon \right)} \right]
F_{\mu \nu} F^{\mu \nu} \; ,
\label{dilatontoWkinetic}
 \end{equation}
where the parameters $\alpha_{W, n}$ are dimensionless. They are 
expected to be of order one. The physical gauge coupling is now given by
 \begin{equation}
\frac{1}{g^2} = \frac{1}{\hat{g}^2}
 \left[ 1 + \sum_{n =1}^{\infty}
\alpha_{W, n} \overline{\lambda}_{\mathcal{O}}^n \;
f^{\left( - n \epsilon \right)} \right] \; .
 \end{equation}
Expanding Eq.~({\ref{dilatontoWkinetic}}) to leading order in $\sigma$ 
we obtain
 \begin{equation}
\epsilon
\frac{\bar{c}_{W}}{4g^2} \frac{\sigma}{f} F_{\mu \nu} F^{\mu \nu} \; ,
 \end{equation}
where the dimensionless parameter $\bar{c}_W$ is given by
 \begin{equation}
\bar{c}_W = \frac{
\sum_{n =1}^{\infty} n \alpha_{W, n} \overline{\lambda}_{\mathcal{O}}^n \;
f^{\left(- n \epsilon \right)}}
{ 1 +  \sum_{n =1}^{\infty} \alpha_{W, n}
\overline{\lambda}_{\mathcal{O}}^n \;
f^{\left(- n \epsilon \right)}} \; .
 \end{equation}
In a strongly coupled theory $\bar{c}_W$ is expected to be of order 
$\overline{\lambda}_{\mathcal{O}} f^{- \epsilon}$, which is the value of 
$\hat{\lambda}_{\mathcal{O}}$ at the breaking scale $f$. It follows that 
this correction to the dilaton coupling is suppressed by $\epsilon 
\hat{\lambda}_{\mathcal{O}}$, which is of order 
$m_{\sigma}^2/\Lambda^2$.

Conformal symmetry violating effects also modify the gauge boson mass
term, Eq.~(\ref{dilatontoW}), which now becomes
 \begin{equation}
\left( \frac{\chi}{f} \right)^2
 \left[ 1 + \sum_{n =1}^{\infty}
\beta_{W, n} \overline{\lambda}_{\mathcal{O}}^n \;
\chi^{\left( - n \epsilon \right)} \right]
\frac{\hat{m}_W^2}{\hat{g}^2} W_{\mu}^+ W^{\mu -} \; .
\label{Wmasslambda}
 \end{equation}
Here $\hat{m}_{W}$ is the $W$ boson mass in the unperturbed theory, and 
the dimensionless parameters $\beta_{W, n}$ are of order one. Expanding 
this out in terms of $\sigma(x)$, we see that to leading order in 
inverse powers of $f$, the dilaton couples as
 \begin{equation}
\frac{\sigma}{f} \frac{m_W^2}{g^2} \left[ 2 + c_{W} \epsilon
\right] W_{\mu}^+ W^{\mu -} \; .
 \end{equation}
Here $m_W^2$ is again the physical $W$ boson mass, 
 \begin{equation}
m_W^2 = \hat{m}_W^2 \left[ 
\frac{1 + \sum_{n =1}^{\infty} \beta_{W, n} 
\overline{\lambda}_{\mathcal{O}}^n \;
f^{\left(- n \epsilon \right)}}
     {1 + \sum_{n =1}^{\infty} \alpha_{W, n} 
\overline{\lambda}_{\mathcal{O}}^n \;
f^{\left(- n \epsilon \right)}}
 \right] \; ,
 \end{equation}
while the dimensionless parameter $c_W$ is given by
 \begin{equation}
c_W = - \frac{
\sum_{n =1}^{\infty} n \beta_{W, n} \overline{\lambda}_{\mathcal{O}}^n \;
f^{\left(- n \epsilon \right)}}
{ 1 +  \sum_{n =1}^{\infty} \beta_{W, n}
\overline{\lambda}_{\mathcal{O}}^n \;
f^{\left(- n \epsilon \right)}} \; .
 \end{equation}
In the strong coupling limit, $c_{W}$ is expected to be of order 
$\overline{\lambda}_{\mathcal{O}} f^{- \epsilon} \sim 
\hat{\lambda}_{\mathcal{O}}$. We see that the effect of the conformal 
symmetry violating term is to correct the dilaton couplings by order 
$\epsilon \hat{\lambda}_{\mathcal{O}} \sim m_{\sigma}^2/\Lambda^2$. 

If we instead consider the limit when $|\delta 
g(\hat{\lambda}_{\mathcal{O}})|$ is greater than $\epsilon$ at the 
symmetry breaking scale $f$, the analysis is very similar. The only 
significant difference is that $\overline{\lambda}_{\mathcal{O}} \chi^{- 
\epsilon}$ in Eqs.~(\ref{chikinlambda}), (\ref{dilatontoWkinetic}) and 
(\ref{Wmasslambda}) is replaced by 
$\Omega(\overline{\lambda}_{\mathcal{O}}, \chi)$, which in this limit is 
given by
 \begin{equation}
\Omega(\overline{\lambda}_{\mathcal{O}}, \chi) =
\frac{\overline{\lambda}_{\mathcal{O}}}
{1 + c_1 \overline{\lambda}_{\mathcal{O}} {\rm log} \chi} \; .
 \end{equation}

Following exactly the same sequence of steps we find that the 
corrections to the dilaton couplings have the same form, but are now 
suppressed by $c_1 \Omega^2 (\overline{\lambda}_{\mathcal{O}}, f)$ 
rather than $\epsilon \overline{\lambda}_{\mathcal{O}} f^{- \epsilon}$. 
However, this new suppression factor is of order $m_{\sigma}^2/ 
\Lambda^2$, exactly as before. We see that the corrections have the same 
form and are of the same size as in the case $|\delta 
g(\hat{\lambda}_{\mathcal{O}})| < \epsilon$. It is not difficult to 
verify that this result is quite general. Therefore, in the remainder of 
the paper we will limit our analysis to the case $|\delta 
g(\hat{\lambda}_{\mathcal{O}})| < \epsilon $, with the understanding 
that the same general conclusions apply to the case $|\delta 
g(\hat{\lambda}_{\mathcal{O}})| > \epsilon $ as well.
 
Next we turn our attention to the dilaton couplings to the massless 
gauge bosons of the SM, the photon and the gluon. Unlike the $W$ and 
$Z$, the Lagrangian for these particles does not break conformal 
invariance at the classical level, only at the quantum level. At one 
loop the renormalization group equations for the corresponding gauge 
couplings are of the form
 \begin{equation} 
\frac{ {\rm d}}{{\rm d} \; {\rm log}\mu} \frac{1}{g^2} = 
\frac{b_<}{8 \pi^2} \; 
 \end{equation}
where the constant $b_< = -11/3$ for electromagnetism and $+7$ for 
color, at scales above the mass of the top.  This implies that under 
infinitesimal scale transformations $x \rightarrow x' = e^{-\omega} x $, 
the operator $F_{\mu \nu} F^{\mu \nu}$ transforms as $ F_{\mu \nu} 
F^{\mu \nu}(x) \rightarrow F'_{\mu \nu} F'^{\mu \nu}(x')$, where 
 \begin{equation} 
F'_{\mu \nu} F'^{\mu \nu}(x') = 
e^{4 \omega} \left( 1 + \frac{b_<}{8 \pi^2} g^2 \omega \right) 
F_{\mu \nu} F^{\mu \nu}(x) 
\label{transform}
 \end{equation}
If conformal symmetry is to be realized nonlinearly, the couplings of 
the dilaton must be such as to compensate for this. It is then easy to 
see that the dilaton couplings in the Lagrangian must take the form
 \begin{equation}
\frac{b_<}{32 \pi^2} {\rm log} \left( \frac{\chi}{f} \right)
F_{\mu \nu} F^{\mu \nu} \; .
 \end{equation} 
Expanding this out in terms of $\sigma(x)$, to leading order in inverse powers
of $f$, we find for the dilaton coupling
 \begin{equation}
\frac{b_<}{32 \pi^2} \frac{\sigma}{f} F_{\mu \nu} F^{\mu \nu} \; .
\label{chitoA}
 \end{equation}
It follows from this that the dilaton couples much more weakly to the 
massless gauge bosons than to the $W$ or the $Z$. The reason is that the 
gauge interactions correspond to marginal operators in the low energy 
effective theory, while mass terms for the gauge bosons are relevant 
operators. Since the dilaton couples as a conformal compensator, it is 
to be expected that its couplings to massless gauge bosons are 
suppressed.

We now consider corrections to this interaction arising from conformal 
symmetry violating effects. These allow direct couplings of the 
compensator $\chi$ to the gauge kinetic term of the form
 \begin{equation}
-\frac{1}{4 \hat{g}^2}  \left[ 1 + \sum_{n =1}^{\infty}
\alpha_{A, n} \overline{\lambda}_{\mathcal{O}}^n \;
\chi^{\left( - n \epsilon \right)} \right]
F_{\mu \nu} F^{\mu \nu} \; .
\label{dilatontoA}
 \end{equation}
The physical gauge coupling is now given by
 \begin{equation}
\frac{1}{g^2} = \frac{1}{\hat{g}^2}
 \left[ 1 + \sum_{n =1}^{\infty}
\alpha_{A, n} \overline{\lambda}_{\mathcal{O}}^n \;
f^{\left( - n \epsilon \right)} \right] \; .
 \end{equation}
Expanding Eq.~({\ref{dilatontoA}}) in terms of $\sigma$, and combining
with Eq.~(\ref{chitoA}) we find for
the dilaton coupling to massless gauge bosons
 \begin{equation}
\frac{\sigma}{f} \left[\frac{b_<}{32 \pi^2}  + \frac{c_{A}}{4g^2} \epsilon
\right] F_{\mu \nu} F^{\mu \nu} \; .
 \end{equation}
Here the dimensionless coupling $c_A$ is given by
 \begin{equation}
c_A = \frac{
\sum_{n =1}^{\infty} n \alpha_{A, n} \overline{\lambda}_{\mathcal{O}}^n \;
f^{\left(- n \epsilon \right)}}
{ 1 +  \sum_{n =1}^{\infty} \alpha_{A, n} 
\overline{\lambda}_{\mathcal{O}}^n \;
f^{\left(- n \epsilon \right)}} \; .
 \end{equation} 
In a strongly coupled theory the parameter $c_A$ is expected to be of 
order $\overline{\lambda}_{\mathcal{O}} f^{- \epsilon} \sim 
\hat{\lambda}_{\mathcal{O}}$ at the scale $f$. We see from this that the 
corrections to the dilaton coupling arising from symmetry breaking 
effects are suppressed by $m_{\sigma}^2/\Lambda^2$. Nevertheless, the 
fact that the leading order effect is loop suppressed and therefore 
small implies that the symmetry breaking contribution may dominate.

Finally we consider the couplings of the dilaton to the SM fermions. In
the limit that conformal symmetry is exact, the coupling of the dilaton
is such as to compensate for the spontaneous breaking of conformal
invariance by the fermion mass terms. These interactions take the form
 \begin{equation}
\frac{\chi}{f} m_{\psi} \bar{\psi} \psi \;
\label{chitopsi}
 \end{equation}
in the potential, where we have suppressed flavor indices. Expanding the 
compensator out in terms of $\sigma$ we obtain
 \begin{equation}
\sigma \frac{m_{\psi}}{f} \bar{\psi} \psi \; .
\label{compquark2dil}
 \end{equation}

However, if a conformal symmetry breaking effect of the form considered
in the previous section is present, Eq.~(\ref{chitopsi}) generalizes to
 \begin{equation}
\label{fmt1}
\frac{\chi}{f} \left[ 1 + \sum_{n =1}^{\infty}
\beta_{\psi, n} \overline{\lambda}_{\mathcal{O}}^n \;
\chi^{\left(- n \epsilon \right)} \right]
\hat{m}_{\psi} \bar{\psi} \psi  \; ,
 \end{equation}
where $\hat{m}_{\psi}$ is the fermion mass in the unperturbed theory, 
and the parameters $\beta_{\psi, n}$ are dimensionless. In obtaining 
this we have assumed that the operator ${\mathcal{O}}$ does not violate 
the approximate U(3$)^5$ flavor symmetry associated with the SM fermions 
in the chiral limit, which is broken by the fermion mass terms. This 
ensures that the dilaton couples diagonally in the mass basis.

In general the operator $\mathcal{O}$ will also correct the fermion kinetic
term, which generalizes to
 \begin{equation}
\label{fkt1}
\left[ 1 + \sum_{n =1}^{\infty}
\alpha_{\psi, n} \overline{\lambda}_{\mathcal{O}}^n \;
\chi^{\left(- n \epsilon \right)} \right]
\overline{\psi} \gamma^{\mu} \partial_{\mu} \psi \; .
 \end{equation}
After expanding out Eqs.~(\ref{fmt1}) and (\ref{fkt1}) in terms of
$\sigma$, rescaling to make the fermion kinetic term canonical, and then
using the equation of motion for $\psi$, we obtain a correction to the
dilaton coupling of the form
 \begin{equation}
\sigma \frac{m_{\psi}}{f} \left[ 1 + c_{\psi} \epsilon
\right]
\bar{\psi} \psi \; ,
 \end{equation}
In a strongly coupled conformal field theory we expect that $c_{\psi}$ 
is of order $\overline{\lambda}_{\mathcal{O}} f^{- \epsilon}$. We 
conclude from this that the effect of the conformal symmetry violating 
terms is to modify the dilaton couplings to the SM fermions by order $ 
m_{\sigma}^2/\Lambda^2$.

From this discussion we see that conformal symmetry violating effects 
associated with the operator ${\mathcal{O}}$ correct the parameters in 
the low energy effective theory at order $\hat{\lambda}_{\mathcal{O}}$. 
However, these effects can be absorbed into the masses and couplings of 
the light states, so that corrections to the form of the dilaton 
couplings to the SM only arise at order $\epsilon 
\hat{\lambda}_{\mathcal{O}} \sim m_{\sigma}^2/\Lambda^2$, and are 
therefore small if the dilaton is light. To understand why the 
corrections to the form of the dilaton couplings receive additional 
suppression, note that if $g(\hat{\lambda}_{\mathcal{O}})$ were to 
vanish close to the breaking scale $f$, the effects of explicit 
conformal symmetry violation would disappear even though $ 
\hat{\lambda}_{\mathcal{O}}$ was non-zero. In this limit, the dilaton 
couplings must have exactly the same form as in a theory without 
conformal symmetry violation, so that all the corrections of order 
$\hat{\lambda}_{\mathcal{O}}$ must be able to be absorbed into the 
masses and couplings of the SM states. It follows that corrections to 
the form of the dilaton couplings must be be suppressed by both 
$\hat{\lambda}_{\mathcal{O}}$ and $g(\hat{\lambda}_{\mathcal{O}})$ 
($\sim \epsilon$ in the limit we are working in).

In summary we see that corrections to the form of the dilaton couplings 
to SM states arising from conformal symmetry violating effects are 
suppressed by the square of the ratio of the dilaton mass to the strong 
coupling scale, and therefore under good theoretical control in the 
theories of interest. These contributions are generally subleading, 
except in the case of dilaton couplings to marginal operators, when 
symmetry violating effects can dominate.

\section{Technicolor}

In this section we determine the form of the couplings of a light 
dilaton to the SM fields in a scenario where electroweak symmetry is 
broken dynamically by a strongly interacting sector, and there is no 
light Higgs. The strongly interacting sector is assumed to be conformal 
in the far ultraviolet. However, conformal symmetry is explicitly broken 
by the operator $\mathcal{O}$, which grows large close to the TeV scale 
triggering electroweak symmetry breaking. The SM gauge fields do not 
constitute part of the strongly interacting sector. However, this sector 
transforms under the weak and electromagnetic gauge interactions. It may 
also transform under the SM color group.  The SM gauge interactions 
constitute another small explicit breaking of the conformal symmetry. The 
SM fermions may be elementary, or may emerge as composites or partial 
composites of the strong dynamics.

The AdS/CFT correspondence relates this class of theories to Higgsless 
Randall-Sundrum models with the SM gauge fields propagating in the bulk.  
The couplings of the radion to SM fields in this framework have been 
determined, in the limit that effects associated with the dynamics that 
stabilizes the radion are neglected~\cite{Rizzo:2002pq, arXiv:0705.3844}. 
We reproduce these results, and in doing so establish their validity 
beyond the large $N$ limit. We also determine the corrections to the 
dilaton couplings that arise from conformal symmetry violating effects.

In order to avoid large corrections to precision electroweak 
observables, the strongly interacting sector must respect a custodial 
SU(2) symmetry. This symmetry is not exact, but is broken by the SM 
Yukawa couplings, and also by hypercharge. A simple way to realize 
custodial symmetry is to extend the SU(2$)_{\rm L}$ symmetry of the SM 
to SU(2$)_{\rm L} \times$ SU(2$)_{\rm R}$. Only the diagonal generator 
of this new SU(2$)_{\rm R}$ is gauged, and is associated with 
hypercharge. The strong dynamics breaks this extended symmetry down to 
the diagonal SU(2), which is identified with the custodial symmetry. 
Only a U(1) subgroup of the original SU(2$) \times$ U(1) gauge symmetry 
survives, and is identified with electromagnetism.

The NGBs $\pi(x)$ that arise from the breaking of SU(2$) 
\times$ SU(2) gauge symmetry down to the custodial SU(2) can be 
parametrized in terms of a matrix $\Sigma$, defined as
 \begin{equation}
\Sigma = e^{i \frac{\pi(x)}{v}}
\left( \begin{array}{cc}
v & 0 \\
0 & v \end{array} \right) \; .
 \end{equation}
Here $v$ is the electroweak VEV. $\Sigma$ transforms linearly under 
SU(2$)_{\rm L} \times$ SU(2$)_{\rm R}$, and is therefore more convenient 
for writing interactions. In unitary gauge the NGBs $\pi(x)$ 
are absorbed into the $W$ and $Z$ gauge bosons and $\Sigma$ can be 
replaced by its VEV.

\subsection{Couplings to Gauge Bosons}

We begin by determining the dilaton couplings to the $W$ and $Z$ gauge 
bosons. In the ultraviolet, the SM gauge interactions do not violate the 
conformal symmetry of the theory at the classical level, only at the 
quantum level. Therefore, when these effects are included, the theory 
still respects conformal symmetry up to effects which are suppressed by 
loops involving the SM gauge bosons. Therefore, the dominant 
interactions of the dilaton to the $W$ and $Z$ bosons in the low energy 
effective theory arise from couplings which compensate for the breaking 
of conformal invariance by the gauge boson mass terms. These take 
exactly the same form in the Lagrangian as in the case of composite $W$ 
and $Z$ gauge bosons
 \begin{equation}
\left( \frac{\chi}{f} \right)^2
\frac{m_W^2}{g^2} W_{\mu}^+ W^{\mu -} \; .
\label{trivialW}
 \end{equation}
Expanding this out in terms of $\sigma$ to leading 
order in inverse powers of $f$, we again find
 \begin{equation}
2 \frac{\sigma}{f} \frac{m_W^2}{g^2} W_{\mu}^+ W^{\mu -} \; .
 \end{equation}
This agrees with the known results for the coupling of the radion to 
bulk gauge bosons in Randall-Sundrum models~\cite{Rizzo:2002pq, 
arXiv:0705.3844}. Our analysis shows that this formula is valid beyond
the large $N$ limit.

When conformal symmetry violating effects associated with the operator 
$\mathcal{O}$ are present, the gauge boson mass will in general depend 
on $\hat{\lambda}_{\mathcal{O}}$. Then Eq.~{\ref{trivialW}) generalizes
to
 \begin{equation}
\left( \frac{\chi}{f} \right)^2
 \left[ 1 + \sum_{n =1}^{\infty}
\beta_{W, n} \overline{\lambda}_{\mathcal{O}}^n \;
\chi^{\left( - n \epsilon \right)} \right]
\frac{\hat{m}_W^2}{\hat{g}^2} W_{\mu}^+ W^{\mu -} \; .
 \end{equation}
Here $\hat{m}_{W}$ is the $W$ boson mass in the unperturbed theory, and
the dimensionless parameters $\beta_{W, n}$ are of order one. Expanding
this out in terms of $\sigma(x)$, we find that the dilaton couples as
 \begin{equation}
\frac{\sigma}{f} \frac{m_W^2}{g^2} \left[ 2 + c_{W} \epsilon
\right] W_{\mu}^+ W^{\mu -} \; ,
 \end{equation}
where $m_W^2$ is the physical $W$ boson mass. The dimensionless 
parameter $c_W$ is of order $\overline{\lambda}_{\mathcal{O}} f^{- 
\epsilon} \sim \hat{\lambda}_{\mathcal{O}}$ at the scale $f$, so that 
the correction to the coupling is suppressed by 
$m_{\sigma}^2/\Lambda^2$.
 
We move on to consider the dilaton couplings to the massless gauge 
bosons of the SM, the gluon and the photon. We first determine the form 
of the couplings in the limit that effects arising from the operator 
$\mathcal{O}$ are neglected. Above the breaking scale, the 
renormalization group equation for the corresponding gauge coupling 
takes the form
 \begin{equation}
\frac{ {\rm d}}{{\rm d} \; {\rm log}\mu} \frac{1}{g_{UV}^2} =
\frac{b_>}{8 \pi^2} \; ,
\label{RGabove}
 \end{equation} 
where the constant $b_>$ receives contributions from both elementary 
states and the strongly interacting sector. Similarly, below the 
breaking scale it takes the form
 \begin{equation}
\frac{ {\rm d}}{{\rm d} \; {\rm log}\mu} \frac{1}{g_{IR}^2} =
\frac{b_<}{8 \pi^2} \;
\label{gIRrunning}
 \end{equation} 
where $b_<$ receives contributions from elementary states, and also from 
any additional light states that emerge from the strongly interacting 
sector after symmetry breaking.
 
Equation~(\ref{RGabove}) indicates that above the symmetry breaking 
scale, under infinitesimal scale transformations $x \rightarrow x' = 
e^{-\omega} x $, the operator corresponding to the gauge kinetic term 
transforms as $ F_{\mu \nu} F^{\mu \nu}(x) \rightarrow F'_{\mu 
\nu} F'^{\mu \nu}(x')$, where
 \begin{equation}
F'_{\mu \nu} F'^{\mu \nu}(x') =
e^{4 \omega} \left( 1 + \frac{b_>}{8 \pi^2} g_{UV}^2 \omega \right)
F_{\mu \nu} F^{\mu \nu}(x)
\label{transform>}
 \end{equation}
Below the symmetry breaking scale the corresponding transformation is $ 
F_{\mu \nu} F^{\mu \nu}(x) \rightarrow F'_{\mu \nu} F'^{\mu 
\nu}(x')$, where
 \begin{equation}
F'_{\mu \nu} F'^{\mu \nu}(x') =
e^{4 \omega} \left( 1 + \frac{b_<}{8 \pi^2} g_{IR}^2 \omega \right)
F_{\mu \nu} F^{\mu \nu}(x)
\label{transform<}
 \end{equation}
Above the symmetry breaking scale we can make the gauge kinetic term 
formally invariant under infinitesimal scale transformations by promoting 
the gauge coupling constant $g_{UV}$ to a spurion that under $x 
\rightarrow x' = e^{-\omega} x$ transforms as
 \begin{equation}
\frac{1}{g_{UV}^2} \rightarrow \frac{1}{{g'}_{UV}^2} =
\frac{1}{g_{UV}^2} - \frac{b_>}{8 \pi^2} \omega \; .
\label{gUVtransform}
 \end{equation}

Now, matching at one loop across the symmetry breaking threshold we have
 \begin{equation}
\frac{1}{g_{IR}^2} = \frac{1}{g_{UV}^2} + \frac{C}{8 \pi^2} \;
\label{gmatch} 
 \end{equation}
where $C$ is a dimensionless number that depends on the gauge quantum 
numbers of the states in the strongly interacting sector that have been 
integrated out at the threshold. While $C$ cannot be calculated, since 
it depends on details of the strong dynamics, it is of order the number 
of states that have masses at the threshold. It is independent of $g^2$
up to corrections which are additionally loop suppressed.
 
In the limit that conformal symmetry violating effects arising from the 
operator $\mathcal{O}$ are neglected, it follows from 
Eqs.~(\ref{transform<}) and (\ref{gUVtransform}) that if the gauge 
kinetic term in the low energy effective theory,
 \begin{equation}
-\frac{1}{4} \frac{1}{g_{IR}^2} F_{\mu \nu} F^{\mu \nu} = 
-\frac{1}{4}
\left\{\frac{1}{g_{UV}^2} + \frac{C}{8 \pi^2}\right\} F_{\mu \nu} F^{\mu \nu}
\; ,
\label{CLagrangian}
 \end{equation}
is to be invariant under infinitesimal scale transformations, the 
conformal compensator must couple as
 \begin{equation}
\frac{\left(b_< - b_> \right)}{32 \pi^2} 
{\rm log} \left(\frac{\chi}{f}\right) F_{\mu \nu} F^{\mu \nu} \; 
 \end{equation}
in the Lagrangian. Expanding this out in terms of $\sigma(x)$, to 
leading order in inverse powers of $f$, we find for the dilaton coupling
 \begin{equation}
\frac{\left(b_< - b_> \right)}{32 \pi^2} \frac{\sigma}{f} 
F_{\mu \nu} F^{\mu \nu} \; .
\label{necessity}
 \end{equation}
This agrees with the result in the literature for the coupling of the 
radion to massless bulk gauge bosons in Randall-Sundrum 
models~\cite{Rizzo:2002pq, arXiv:0705.3844}. Our analysis establishes 
that this result is valid beyond the large $N$ limit. This formula is 
valid at scales slightly below the strong coupling scale $4 \pi f$, and 
must be renormalization group evolved to the dilaton mass. If the 
conformal sector does not transform under the SM color group, as may be 
the case in theories where the top quark is not a composite of the 
strong dynamics, then $b_< = b_>$ and the gluon does not couple to the 
dilaton at this order. In such a scenario, the leading interaction of 
the dilaton with the gluons is through a loop of top quarks, just as for 
the Higgs in the SM.

When conformal symmetry violating effects arising from the operator 
$\mathcal{O}$ are included, this formula will receive corrections. The 
renormalization group evolution of the gauge coupling above the symmetry 
breaking scale is affected by the presence of the deformation, with the 
result that $b_>$ is now a function of $\hat{\lambda}_{\mathcal{O}}$, 
$b_> = b_> (\hat{\lambda}_{\mathcal{O}})$. However, the theory remains 
formally invariant under infinitesimal scale transformations if $g_{UV}$ 
is promoted to a spurion as in Eq.~(\ref{gUVtransform}). Hence this 
effect does not alter the form of Eq.~(\ref{necessity}). The operator 
$\mathcal{O}$ also affects the low energy theory through the fact that 
the value of the gauge coupling at low energies depends on the detailed 
spectrum of states at the threshold, which in turn depends on 
$\hat{\lambda}_{\mathcal{O}}$. As a consequence the constant $C$ in 
Eq.~({\ref{gmatch}}) is in general a function of 
$\hat{\lambda}_{\mathcal{O}}$. Now, conformal symmetry ensures that in 
the low energy effective theory $C$ depends on 
$\overline{\lambda}_{\mathcal{O}}$ in the specific combination 
$C(\overline{\lambda}_{\mathcal{O}} \chi^{-\epsilon})$. Since the 
Lagrangian is limited to terms with positive integer powers of 
$\hat{\lambda}_{\mathcal{O}}$, we can expand $C$ as
 \begin{equation}
C = \left[ C_0 + \sum_{n =1}^{\infty}
C_{n} \overline{\lambda}_{\mathcal{O}}^n \;
\chi^{\left(- n \epsilon \right)} \right] \; .
 \end{equation}
Inserting this into Eq.~(\ref{CLagrangian}) and requiring invariance under
(spurious) scale transformations,
we find that the dilaton couplings must take the form
 \begin{equation}
\frac{\sigma}{f} \left[\frac{\left(b_< - b_> \right)}{32 \pi^2}  
+ \frac{\bar{c}_{A}}{ 32 \pi^2} \epsilon
\overline{\lambda}_{\mathcal{O}} f^{-\epsilon} \right] 
F_{\mu \nu} F^{\mu \nu} \; .
\label{diltophotonandgluon}
 \end{equation} 
Here the dimensionless constant $\bar{c}_A$ is expected to be of order 
the number of states that transform under the gauge symmetry that have 
masses at the threshold, so that $\bar{c}_A \sim (b_< - b_>)$. We can 
therefore rewrite Eq.~(\ref{diltophotonandgluon}) as
 \begin{equation}
\frac{\sigma}{f} \frac{\left(b_< - b_> \right)}{32 \pi^2}
\left[1 + c_{A}\epsilon \right]
F_{\mu \nu} F^{\mu \nu} \; ,
\label{diltophotonandgluon2}
 \end{equation} 
 where the dimensionless constant $c_A$ is of order 
$\overline{\lambda}_{\mathcal{O}} f^{-\epsilon}$, which is the value of 
$\hat{\lambda}_{\mathcal{O}}$ at the symmetry breaking scale $f$. Here 
$b_> (\hat{\lambda}_{\mathcal{O}})$ is to be evaluated close to the 
breaking scale. We see from this that in this scenario, corrections to 
the form of the dilaton couplings to massless gauge bosons arising from 
conformal symmetry violating effects are subleading, being suppressed 
not just by $m_{\sigma}^2/\Lambda^2$, but also by a loop factor. This is 
in contrast to the case of composite gauge bosons considered in the 
previous section.

\subsection{Couplings to Fermions} 

\subsubsection{Elementary Fermions}

Next we consider the dilaton couplings to the SM fermions, which we 
label by $Q, U^c, D^c, L$ and $E^c$. These depend on how the fermion 
masses are generated. One possibility is that the fermion masses arise 
from a contact term that couples a scalar operator $\mathcal{H}$ in the 
conformal field theory that carries the gauge quantum numbers of the SM 
Higgs to elementary fermions. For the up-type quarks, this takes the form
 \begin{equation}
y^{ij} \mathcal{H} {Q}_i U^c_j + \; {\rm h.c.}  
\label{techniyukawa}
 \end{equation} 
in the Lagrangian. Here $i$ and $j$ are flavor indices. This leads to
a mass term 
 \begin{equation}
m^{ij} {Q}_i U^c_j  + \; {\rm h.c.} \;
 \end{equation}
in the potential of the low energy effective theory. The generalization 
to the down-type quarks and leptons is straightforward. In the limit 
that $y^{ij}$ is set to zero the ultraviolet theory has a U(3$)_Q \times 
$ U(3$)_U$ flavor symmetry, which can be restored by promoting $y^{ij}$ 
to a spurion that transforms as an anti-fundamental under each of these 
symmetries. By requiring that the low energy effective theory be 
invariant under this spurious flavor symmetry, it follows that $m^{ij}$ 
is proportional to $y^{ij}$ to lowest order in the couplings $y$.

It has been shown that the flavor problem and the large mass of the top 
quark can both be addressed in this framework if the operator 
$\mathcal{H}$ has dimension $\Delta_{\mathcal{H}} \lesssim 1.3$. 
However, if the hierarchy problem is to be solved, the dimension 
$\Delta_{\mathcal{H}^{\dagger} \mathcal{H}}$ of the operator 
$\mathcal{H}^{\dagger} \mathcal{H}$ must satisfy 
$\Delta_{\mathcal{H}^{\dagger} \mathcal{H}} \gtrsim 
4$~\cite{hep-ph/0409274}. Determining whether scalar operators that 
satisfy these criteria can exist in unitary, causal conformal field 
theories is an open question that has attracted considerable recent 
interest~\cite{Rattazzi:2008pe, Vichi:2011ux, Poland:2011ey}, see also 
\cite{Poland:2010wg}. Note that this condition cannot be satisfied in 
the large $N$ limit, and therefore realistic models of this type cannot 
be constructed within the Randall-Sundrum framework.

In order to determine the coupling of the dilaton to the up-type quarks, 
we make the coupling in Eq.~(\ref{techniyukawa}) formally invariant 
under scale transformations by promoting $y^{ij}$ to a spurion that 
transforms as $y^{ij} \rightarrow y'^{ij} = e^{\omega \left(1 - 
\Delta_{\mathcal{H}} \right)}y^{ij}$ under $x \rightarrow x' = 
e^{-\omega} x $. Then the coupling of the dilaton to the up-type quarks 
in the effective theory must respect this symmetry. Since the quark mass 
matrix $m^{ij}$ is proportional to $y^{ij}$ the conformal compensator 
couples as
 \begin{equation}  
m^{ij} \left( \frac{\chi}{f} \right)^{\Delta_{\mathcal{H}}} {Q}_i U^c_j
 + \; {\rm h.c.}
 \label{necessary}
 \end{equation} 
Then to lowest order in inverse powers of $f$, the dilaton coupling to
up-type quarks takes the form~\cite{arXiv:1002.1721}
 \begin{equation}
m^{ij} \Delta_{\mathcal{H}}
\left(\frac{\sigma}{f} \right){Q}_i U^c_j  + \; {\rm h.c.}
\label{dil2quarks}
 \end{equation}
 We see that to the extent that $\Delta_{\mathcal{H}}$ differs from one, 
the dilaton couplings to fermions can differ significantly from those of 
a SM Higgs. 

Once effects of the operator ${\mathcal{O}}$ are included the scaling 
dimension of the operator ${\mathcal{H}}$ receives corrections, 
$\Delta_{\mathcal{H}} = \Delta_{\mathcal{H}}({\hat{\lambda}}_{\mathcal{O}})$. 
However, above the breaking scale the theory continues to remain 
invariant under the infinitesimal spurious scale transformation $y^{ij} \rightarrow 
y'^{ij} = e^{\omega \left(1 - \Delta_{\mathcal{H}} \right)}y^{ij}$ when 
$x \rightarrow x' = e^{-\omega} x $, and so the form of 
Eq.~(\ref{dil2quarks}) is not affected by this. Instead, the leading 
correction arises from the fact that in addition to the term in 
Eq.~(\ref{necessary}), other terms involving the invariant 
$\overline{\lambda}_{\mathcal{O}} \chi^{-\epsilon}$ can now also 
contribute. As a result, Eq.~(\ref{dil2quarks}) is modified to
 \begin{equation}
 m^{ij}
\left(\Delta_{\mathcal{H}} + c_q \epsilon \right)
\frac{\sigma}{f} {Q}_i U^c_j  + \; {\rm h.c.} \; ,
 \end{equation}
 where the dimensionless parameter $c_q$ is of order 
$\overline{\lambda}_{\mathcal{O}} f^{-\epsilon}$. Here 
$\Delta_{\mathcal{H}}({\hat{\lambda}}_{\mathcal{O}})$ is to be evaluated close 
to the symmetry breaking scale. We see that corrections to the form of 
Eq.~(\ref{dil2quarks}) from conformal symmetry violating effects are of 
order $m_{\sigma}^2/\Lambda^2$, and under control.

More generally, there could be several scalar operators 
$\mathcal{H}_{\alpha}$ in the conformal field theory that couple to the 
SM fermions. The coupling in Eq.~(\ref{techniyukawa}) then generalizes 
to
 \begin{equation}
y^{\alpha ij} \mathcal{H}_{\alpha} {Q}_i U^c_j  + \; {\rm h.c.} \; ,
\label{techniflavoryukawa}
 \end{equation}
where the index $\alpha$ runs over all the scalar operators in the 
theory with the quantum numbers of the SM Higgs. However, operators
with dimension significantly larger than one are not expected to play
a significant role.

It follows from the U(3$)_Q \times $ U(3$)_U$ flavor symmetry that 
the up-type fermion masses depend on the couplings $y^{\alpha ij}$ as 
 \begin{equation}
m^{ij} = y^{\alpha ij} D_{\alpha} \; ,
\label{eq54}
 \end{equation}
where the parameters $D_{\alpha}$ depends on the details of the conformal 
field theory. In order to determine the couplings of the dilaton, we 
make the coupling in Eq.~(\ref{techniflavoryukawa}) formally invariant under 
scale transformations by promoting the $y^{\alpha ij}$ to spurions that 
transform as
 \begin{equation}
y^{\alpha ij} \rightarrow y'^{\alpha i j} = e^{\omega \left(1 - 
\Delta_{\mathcal{H}_{(\alpha)}} \right)}y^{(\alpha) ij} \; ,
 \end{equation}
under $x \rightarrow x' = e^{-\omega} x$. There is no sum over $\alpha$ 
on the right hand side of this equation. The various terms in the sum on 
the right hand side of Eq.~(\ref{eq54}) transform differently under this 
transformation. In order to account for this we define
 \begin{equation}
m^{\alpha ij} = y^{\left(\alpha \right) ij} D_{\left(\alpha \right)}  \; ,
 \end{equation}
where again there is no sum over $\alpha$ on the right hand side of this 
equation. Then the requirement that the fermion mass in the low energy 
effective theory be formally invariant under this symmetry constrains 
the conformal compensator to couple as
 \begin{equation}
m^{\alpha ij} 
\left( \frac{\chi}{f} \right)^{\Delta_{\mathcal{H}_{\alpha}}} 
{Q}_i U^c_j  + \; {\rm h.c.}
 \end{equation}
in the potential. This leads to the dilaton coupling
 \begin{equation}
m^{\alpha ij} \Delta_{\mathcal{H}_{\alpha}} \frac{\sigma}{f} {Q}_i U^c_j 
 + \; {\rm h.c.}
\label{dil2quarks1.12}
 \end{equation}
We see from this that if the $\Delta_{\mathcal{H}_{\alpha}}$ are not all 
equal, the couplings of the dilaton in the low energy effective theory 
violate flavor. However, in the absence of large cancellations among the 
contributions of different operators to the quark masses, the matrix 
$m^{\alpha ij} \Delta_{\mathcal{H}_{\alpha}}$ will be somewhat aligned 
with the quark mass matrix $m^{ij}$, leading to suppression of flavor 
violation.

\subsubsection{Partially Composite Fermions}

Another possible origin for the fermion masses is that the SM quarks and 
leptons are partial composites of the strongly interacting 
sector~\cite{Kaplan:1991dc}. This scenario can arise if the theory 
contains elementary fermions ${Q}_i, U^c_i, D^c_i, {L}_i$ and $E^c_i$ 
with the same gauge quantum numbers as the corresponding SM fermions 
that mix with operators in the conformal field theory. The physical SM 
fermions emerge as a linear combination of the corresponding elementary 
particles and states associated with the strongly interacting sector. 
Within the Randall-Sundrum framework, this corresponds to putting the SM 
fermions in the bulk of the space~\cite{Grossman:1999ra}.

To understand this in greater detail, let us consider the mass terms for 
the up-type quarks. These can be generated if the conformal field theory 
contains fermionic operators $\mathcal{Q}^c_{\alpha}$ and 
$\mathcal{U}_{\alpha}$, with dimensions $\Delta_{\mathcal{Q}}$ and 
$\Delta_{\mathcal{U}}$ respectively, that couple to elementary 
fermions ${Q}_i$ and $U^c_i$ in the Lagrangian as
 \begin{equation}
y_{\mathcal{Q}}^{\alpha i} \mathcal{Q}^c_{\alpha} Q_i +
y_{\mathcal{U}}^{\beta j} \mathcal{U}_{\beta} U^c_j 
 + \; {\rm h.c.} 
\label{techniyukawa2}
 \end{equation}
We assume that the indices $\alpha$ and $\beta$, which run from 1 to 
3, are associated with an internal U(3) symmetry of the 
conformal sector so that $\Delta_{\mathcal{Q}}$ and 
$\Delta_{\mathcal{U}}$ are independent of $\alpha$ and $\beta$. We will 
relax this assumption later. If $\Delta_{\mathcal{Q}}$ and 
$\Delta_{\mathcal{U}}$ are close to 5/2, these interactions correspond 
to marginal operators in the conformal field theory. These couplings 
will generate up-type quark masses in the potential of the form
 \begin{equation}
m^{i j} Q_{i} U^c_j  + \; {\rm h.c.}
 \end{equation}
This framework can be extended to the down-type quarks and leptons in a 
straightforward way. The AdS/CFT correspondence relates the operator 
dimensions $\Delta_{\mathcal{Q}}$ and $\Delta_{\mathcal{U}}$ to the mass 
terms for bulk fermions in Randall-Sundrum models.

In the limit that the couplings $y_{\mathcal{Q}}$ and $y_{\mathcal{U}}$ 
are set to zero the ultraviolet theory has a U(3$)_Q \times $ U(3$)_U$ 
flavor symmetry. This symmetry can be restored by promoting 
$y_{\mathcal{Q}}$ and $y_{\mathcal{U}}$ to spurions that transform as 
anti-fundamentals under U(3$)_Q$ and U(3$)_U$ respectively. Then, 
requiring the low energy effective theory to respect this spurious 
symmetry constrains the mass matrix to be proportional to the product 
of $y_{\mathcal{U}}$ and $y_{\mathcal{Q}}$,
 \begin{equation}
m^{ij} \propto \left[ y_{\mathcal{Q}}^T \; y_{\mathcal{U}} \right]^{ij}
 \; ,
 \end{equation}
to lowest order in the couplings $y$. The kinetic terms of the quarks in 
the low energy effective theory also receive corrections from the 
couplings $y_{\mathcal{Q}}$ and $y_{\mathcal{U}}$ of the form
 \begin{equation}
\Delta Z_Q \overline{Q} {\gamma}^{\mu} D_{\mu} Q  +
\Delta Z_U \overline{U}^c {\gamma}^{\mu} D_{\mu} U^c
\label{kineticdil}
 \end{equation}
where
 \begin{eqnarray}
\Delta Z_Q &\sim& \frac{1}{16 \pi^2}
\frac{ y_{\mathcal{Q}}^{\dagger} y_{\mathcal{Q}}}
{f^{5 - 2 \Delta_{\mathcal{Q}}}}  \nonumber \\
\Delta Z_U &\sim& \frac{1}{16 \pi^2}
\frac{ y_{\mathcal{U}}^{\dagger} y_{\mathcal{U}}}
{f^{5 - 2 \Delta_{\mathcal{U}}}} \; .
 \end{eqnarray}
The corrections to the kinetic terms are a consequence of the fact that 
the fermions in the low energy theory are partially composite.

In order to determine the coupling of the dilaton to the up-type quarks, 
we promote $y_{\mathcal{Q}}$ and $y_{\mathcal{U}}$ to spurions that 
transform as $y_{\mathcal{Q}} \rightarrow y'_{\mathcal{Q}} = e^{\omega 
\left(5/2 - \Delta_{\mathcal{Q}} \right)}y_{\mathcal{Q}}$ and 
$y_{\mathcal{U}} \rightarrow y'_{\mathcal{U}} = e^{\omega \left(5/2 - 
\Delta_{\mathcal{U}} \right)}y_{\mathcal{U}}$ under $x \rightarrow x' = 
e^{-\omega} x$. Then the couplings~(\ref{techniyukawa2}) are formally 
invariant under scale transformations, and the conformal compensator 
couples to quarks so as to make low energy effective theory consistent 
with this symmetry. To lowest order in powers of $y_{\mathcal{Q}}$ and 
$y_{\mathcal{U}}$, and neglecting effects arising from $\mathcal{O}$, 
this coupling takes the form
 \begin{equation}
m^{ij} Q_{i} U^c_j 
\left( \frac{\chi}{f}
\right)^{\left(\Delta_{\mathcal{U}} + \Delta_{\mathcal{Q}}-4 \right)}
 + \; {\rm h.c.} 
 \end{equation}
in the potential.
This leads to the dilaton couplings
 \begin{equation}
m^{ij} \left(\Delta_{\mathcal{U}} + \Delta_{\mathcal{Q}} - 4 \right)
\frac{\sigma}{f} Q_{i} U^c_j  + \; {\rm h.c.} 
\label{diltocomp}
 \end{equation}
This agrees with the results in the literature for the coupling of the 
dilaton to partially composite fermions in the large $N$ 
limit~\cite{arXiv:1002.1721}, and for the coupling of the radion to bulk 
fermions in the Randall-Sundrum model~~\cite{Rizzo:2002pq, 
arXiv:0705.3844}. Our analysis establishes that these results are valid 
beyond the large $N$ limit. 

When effects of the operator ${\mathcal{O}}$ are included, 
Eq.~(\ref{diltocomp}) is modified to
 \begin{equation}
m^{ij} \left[ \left(\Delta_{\mathcal{U}} + \Delta_{\mathcal{Q}} - 4 \right)
+ c_q \epsilon \right]
\frac{\sigma}{f} Q_{i} U^c_j  + \; {\rm h.c.} \; ,
 \end{equation} 
 where $c_q$ is of order $\overline{\lambda}_{\mathcal{O}} 
f^{-\epsilon}$, which is the value of $\hat{\lambda}_{\mathcal{O}}$ at 
the scale $f$. In this expression, 
$\Delta_{\mathcal{U}}(\hat{\lambda}_{\mathcal{O}})$ and 
$\Delta_{\mathcal{Q}}(\hat{\lambda}_{\mathcal{O}})$ are to be evaluated 
close to the symmetry breaking scale. In obtaining this result, we have 
assumed that the operator $\mathcal{O}$ does not break the approximate 
SM flavor symmetries, or the internal U(3) symmetry of the conformal 
sector. It follows that corrections to the form of Eq.~(\ref{diltocomp}) 
from conformal symmetry violating effects are suppressed by 
$m_{\sigma}^2/\Lambda^2$, and are under good theoretical control.

There are additional contributions to the dilaton coupling to quarks 
associated with the corrections to the kinetic terms, 
Eq.~(\ref{kineticdil}). However, using the equations of motion, it can 
be shown these contributions are higher order in $y_{\mathcal{Q}}$ and 
$y_{\mathcal{U}}$ than the effects we have considered, and are therefore 
suppressed.

In the more general case the operators $\mathcal{Q}^c_{\alpha}$ and 
$\mathcal{U}_{\alpha}$, could have dimensions 
$\Delta_{\mathcal{Q}_{\alpha}}$ and $\Delta_{\mathcal{U}_{\alpha}}$ that 
depend on the flavor index $\alpha$. Then it follows from the spurious 
flavor symmetries that the SM fermion masses depend on the couplings 
$y_{\mathcal{Q}}$ and $y_{\mathcal{U}}$ as
 \begin{equation}
m^{ij} = y_{\mathcal{Q}}^{\alpha i} y_{\mathcal{U}}^{\beta j}
D_{\alpha \beta} \; ,
 \end{equation}
where the parameter $D_{\alpha \beta}$ depends on the details of the 
conformal field theory. We can make the theory formally invariant under 
scale transformations by promoting $y_{\mathcal{Q}}$ and 
$y_{\mathcal{U}}$ to spurions that transform as 
 \begin{eqnarray}
y^{\alpha i}_{\mathcal{Q}} &\rightarrow& y'^{\alpha i}_{\mathcal{Q}} 
= e^{\omega \left(5/2 - \Delta_{\mathcal{Q}_{\left(\alpha \right)}} \right)}
y^{\left(\alpha \right) i}_{\mathcal{Q}} \nonumber \\ 
y^{\alpha i}_{\mathcal{U}} &\rightarrow& y'^{\alpha i}_{\mathcal{U}} 
= e^{\omega \left(5/2 - \Delta_{\mathcal{U}_{\left(\alpha \right)}} \right)} 
y^{\left( \alpha \right) i}_{\mathcal{U}} 
 \end{eqnarray}
under $x \rightarrow x' = e^{-\omega} x^{\mu}$, where there is no sum 
over $\alpha$ on the right hand side of the equations. The couplings of 
the dilaton in the low energy effective theory must respect this 
symmetry. We define
 \begin{equation}
m^{\alpha \beta i j} = y^{(\alpha) i}_{\mathcal{Q}}
y^{(\beta) j}_{\mathcal{U}} D_{(\alpha) (\beta)} \; ,
 \end{equation}
where again there is no sum over $\alpha$ and $\beta$ on the right hand
side of the equation. We also define 
 \begin{equation}
\Delta^{\mathcal{Q} \mathcal{U}}_{\alpha \beta} =
\Delta_{\mathcal{Q}_{\alpha}} +  \Delta_{\mathcal{U}_{\beta}} - 4 \; .
 \end{equation}
In terms of these new variables the coupling of the conformal 
compensator to the up-type quarks can be expressed as
 \begin{equation}
m^{\alpha \beta i j} Q_{i} U^c_j
\left( \frac{\chi}{f}
\right)^{\Delta^{\mathcal{Q} \mathcal{U}}_{\alpha \beta}}
 + \; {\rm h.c.}
 \end{equation}
Expanding this out, we find for the dilaton couplings in the potential
 \begin{equation}
m^{\alpha \beta i j} \Delta^{\mathcal{Q} \mathcal{U}}_{\alpha \beta}
\frac{\sigma}{f} Q_{i} U^c_j  + \; {\rm h.c.}
\label{diltocompquarks}
 \end{equation}
It follows that in this scenario, the couplings of the dilaton to the 
quarks in the low energy effective theory violate flavor. However, in 
the absence of large cancellations among the contributions of the 
different $y_{\mathcal{Q}}$ and $y_{\mathcal{U}}$ to $m^{ij}$, we expect 
some degree of alignment between the quark mass matrix and the dilaton 
coupling matrix, which may be sufficient to satisfy flavor bounds.

Since $\mathcal{Q}^c$ and $\mathcal{U}$ are part of the strongly 
interacting sector, they must arise from complete multiplets of 
SU(2$)_{\rm L} \times$ SU(2$)_{\rm R}$. There are two distinct 
possibilities for the realization of this symmetry, which we consider in 
turn.

The first possibility is that $\mathcal{Q}^c$ transforms as $(2,1)$ 
under SU(2$)_{\rm L} \times$ SU(2$)_{\rm R}$ while $\mathcal{U}$ is 
partnered by another state $\mathcal{D}$, and together they transform as 
a $(1,2)$. In the context of Randall-Sundrum models, this realization of 
custodial symmetry was first proposed in~\cite{Agashe:2003zs}. The 
large mass of the top quark implies that the couplings $y_{\mathcal{Q}}$ 
and $y_{\mathcal{U}}$ must be sizable for the third generation quarks. 
This realization leads to mild tension with precision electroweak tests, 
since $y_{\mathcal{U}}$ distinguishes between $\mathcal{U}$ and 
$\mathcal{D}$, and therefore violates custodial SU(2) symmetry.

The alternative possibility~\cite{Agashe:2006at} is that $\mathcal{Q}^c$ 
is partnered with a new state $\hat{\mathcal{Q}}^c$, and together they 
transform as $(2,2)$ under SU(2$)_{\rm L} \times$ SU(2$)_{\rm R}$. 
Meanwhile, $\mathcal{U}$ is now just a singlet. In this realization of 
the extended symmetry, it is $y_{\mathcal{Q}}$ that violates custodial 
SU(2) and leads to tension with precision tests. This difficulty can be 
avoided if the third generation SU(2) singlet up-type quark $U_3^c$ is a 
composite of the strongly interacting sector. This allows 
$y_{\mathcal{Q}}$ to remain small enough to avoid conflict with the 
bound. In this scenario Eq.~(\ref{diltocompquarks}) remains valid, the 
only difference being that $\Delta_{\mathcal{U}_3}$ now takes the value 
5/2.

\section{Higgs as a pNGB}

Next we consider theories where the SM Higgs doublet emerges as the pNGB 
associated with the breaking of an approximate global symmetry by strong 
conformal dynamics. For concreteness, we will take the global symmetry 
to be SO(6), which is broken to SO(5). An SU(2$) \times $ U(1) subgroup 
of the unbroken SO(5) is gauged, and identified with the electroweak 
gauge sector of the SM. Of the 5 pNGBs, 4 are identified with the SM 
Higgs doublet, while the remaining one is a SM singlet.

For the purpose of writing interactions, it is convenient to work in a 
framework where we keep only the symmetries associated with the SU(3$) 
\times$ U(1) subgroup of the non-linearly realized SO(6) global symmetry 
manifest. As shown in \cite{arXiv:0710.0333}, the 5 NGBs 
associated with the breaking of SO(6) to SO(5) can be identified with 
the 5 NGBs arising from the breaking of the SU(3$) \times$ 
U(1) subgroup of SO(6) to SU(2$) \times$ U(1), since the corresponding 
coset spaces are identical. We parametrize the NGBs as 
$h^a$, where $a$ runs from 1 to 5. Rather than work with the $h^a$ 
directly it is more convenient to construct an object $\phi$ which 
transforms linearly under SU(3$) \times$ U(1).
 \begin{equation}
\phi = \hat{f} {\rm exp} \left( i h^a t^a \right) 
\left( {\begin{array}{c}
 0  \\
 0  \\
 1  \\
 \end{array} } \right)  \; .
\label{phidefn}
 \end{equation}
Note that we are employing a convention where the $h^a$ carry no mass 
dimension. We expect that $\hat{f}$ and $f$ will be of the same order, 
since the same dynamics is responsible for the breaking of both 
conformal symmetry and the global symmetry. The 5 matrices $t^a$ span 
[SU(3$) \times$ U(1)/SU(2$) \times$ U(1)], and are chosen as
 \begin{eqnarray}
\left\{t^a\right\} &=& \left\{ \pmatrix{ 0 & 0 & 0 \cr
                0 & 0 & -i \cr
                0 & i & 0}, \pmatrix{ 0 & 0 & 0 \cr
                0 & 0 & 1 \cr
                0 & 1 & 0},\  \pmatrix{ 0 & 0 & -i \cr
                0 & 0 & 0 \cr
                i & 0 & 0},\ \right.\nonumber \\
&&\left. \pmatrix{ 0 & 0 & 1 \cr
                0 & 0 & 0 \cr
                1 & 0 & 0}, \sqrt{\frac{2}{3}}\pmatrix{ 1 & 0 & 0 \cr
                0 & 1 & 0 \cr
                0 & 0 & -1} \right\} .
 \end{eqnarray}
This choice allows us to take $h^a$, $a = 1 \rightarrow 4$ to represent 
the SM Higgs doublet, which we denote by $h$, while $h^5$ represents the 
additional singlet.

The low energy effective Lagrangian will in general contain all possible
operators consistent with the SU(3$) \times$ U(1) global symmetry, but
with restrictions on the coefficients of various terms enforced by the
larger SO(6) symmetry. In particular the dangerous custodial SU(2)
violating operator
 \begin{equation}
\left| \phi^{\dagger} D_{\mu} \phi - (D_{\mu} \phi)^{\dagger} \phi \right|^2 
\;,
 \end{equation}
while allowed by SU(3$) \times$ U(1), is forbidden by SO(6). Here 
$D_{\mu}$ is the gauge covariant derivative with respect to the SM 
SU(2) $\times$ U(1) gauge symmetry.

The requirement of scale invariance implies that the non-linear sigma 
model condition $|\phi|^2 = \hat{f}^2$ becomes $|\phi|^2 = \hat{f}^2 
\hat{\chi}^2$. This means that we can make the low energy effective 
theory for the pNGBs invariant under scale transformations, up to terms 
arising from effects that explicitly violate the conformal and global 
symmetries, by making the replacement $\hat{f} \rightarrow \hat{f} 
\hat{\chi}$ in Eq.~(\ref{phidefn}). The net effect is that in the low 
energy effective theory the $h^a$ transform as fields with scaling 
dimension equal to zero, up to effects that violate conformal symmetry. 
This allows us to determine the form of the dilaton couplings to the SM 
fields. A major simplification is that since $h$ has no scaling 
dimension, when replaced by its VEV the various operators have exactly 
the same scaling dimensions as in the technicolor models of the previous 
section, and many results can simply be carried over.

\subsection{Couplings to Gauge Bosons}

We begin by considering the dilaton couplings to the weak gauge bosons 
of the SM. These arise from the gauge covariant kinetic term for $\phi$,
 \begin{equation}
\left( D_{\mu} \phi \right)^{\dagger} D^{\mu} \phi  \;.
\label{kineticphi}
 \end{equation}
Expanding out $\phi$ to lowest order in $h$, we obtain the gauge 
covariant kinetic term for the SM Higgs doublet
 \begin{equation}
\hat{\chi}^2 \hat{f}^2 \left( D_{\mu} h \right)^{\dagger} D^{\mu} h  \;,
\label{kinetich}
 \end{equation}
Working in unitary gauge, and replacing $h$ by its VEV, we find the 
coupling of the conformal compensator to the $W$ bosons in the Lagrangian
 \begin{equation}
\frac{m_W^2}{g^2} \frac{\chi^2}{f^2} W_{\mu}^+ W^{\mu \; -} \; .
 \end{equation} 
This leads to the dilaton coupling
 \begin{equation}
2 \frac{\sigma}{f} \frac{m_W^2}{g^2} W_{\mu}^+ W^{\mu \; -} \; .
 \end{equation}
As expected, this is identical to the corresponding formula in the 
technicolor case. When conformal symmetry violating effects arising from 
${\mathcal{O}}$ are incorporated, the non-linear sigma model condition 
is modified to
 \begin{equation}
|\phi|^2 = 
\hat{f}^2 \hat{\chi}^2
 \left[ 1 + \sum_{n =1}^{\infty}
\alpha_{\phi, n} \overline{\lambda}_{\mathcal{O}}^n \;
\chi^{\left( - n \epsilon \right)} \right] \; ,
 \end{equation}
where the dimensionless parameters $\alpha_{\phi, n}$ are expected to be
of order one. Then the dilaton coupling to $W$ bosons is modified to
 \begin{equation}
\frac{\sigma}{f} \frac{m_W^2}{g^2} 
\left( 2 + c_W \epsilon \right) W_{\mu}^+ W^{\mu \; -} \; ,
 \end{equation}
where $c_W$ is of order $\overline{\lambda}_{\mathcal{O}} 
f^{-\epsilon}$. We see that the corrections are suppressed by 
$m_{\sigma}^2/\Lambda^2$, exactly as in the technicolor case.

Next we consider dilaton couplings to the massless gauge bosons of the 
SM, the gluons and the photon. The leading effect which breaks conformal 
invariance is again the running of the gauge couplings, just as in the 
previous section. The results can simply be carried over, and are given 
by Eq.~(\ref{diltophotonandgluon2}),
 \begin{equation}
\frac{\sigma}{f} \frac{\left(b_< - b_> \right)}{32 \pi^2}
\left[1 + c_{A}\epsilon \right]
F_{\mu \nu} F^{\mu \nu} \; .
\label{diltophoton2}
 \end{equation}
 Here $b_> (\hat{\lambda}_{\mathcal{O}})$ is to be evaluated close to 
the breaking scale. As can be seen from this formula, corrections to the 
form of the dilaton couplings from conformal symmetry breaking effects 
are suppressed by $m_{\sigma}^2/\Lambda^2$ and also by a loop factor, 
and are generally small.

\subsection{Couplings to Fermions}

\subsubsection{Elementary Fermions}

Next we consider dilaton couplings to the SM fermions. We begin with the 
case where the SM fermions are elementary, and their masses arise from 
direct contact interactions with operators in the conformal field theory. 
The up-type fermion masses arise from terms in the Lagrangian of the form
 \begin{equation}
\hat{y}^{ij} \mathcal{H} {Q}_i U^c_j  + \; {\rm h.c.} 
\label{techniyukawa2.1}
 \end{equation}
that break the global symmetry. Here the operator $\mathcal{H}$ has the 
quantum numbers of the SM Higgs doublet. This leads to Yukawa couplings 
for the up-type quarks in the potential of the low energy effective 
theory,
 \begin{equation}
{y}^{ij} \left( \hat{f}{h} \right) {Q}_i U^c_j  + \; {\rm h.c.} \; , 
 \end{equation}
where we are neglecting higher order terms in $h$ which may also arise 
from the term in Eq.~(\ref{techniyukawa2.1}). It follows from the flavor 
symmetries that ${y}^{ij}$ is proportional to $\hat{y}^{ij}$ to lowest 
order in the couplings $\hat{y}$. We can find the dilaton couplings by 
promoting $\hat{y}^{ij}$ to a spurion exactly as in the technicolor 
case. Noting that $h$ has no scaling dimension, it follows that the 
conformal compensator couples to up-type quarks as
 \begin{equation}
y^{ij} \left( \frac{\chi}{f} \right)^{\Delta_{\mathcal{H}}}
\left( \hat{f} h \right) {Q}_i U^c_j  + \; {\rm h.c.} 
 \end{equation}
in the potential.
This leads to the dilaton coupling
 \begin{equation}
y^{ij} \Delta_{\mathcal{H}}
\frac{\sigma}{f} \left( \hat{f} h \right) {Q}_i U^c_j  + \; {\rm h.c.} \; .
\label{dil2quarksh}
 \end{equation}
Replacing $h$ by its VEV we obtain
 \begin{equation} 
m^{ij} \Delta_{\mathcal{H}} \frac{\sigma}{f} {Q}_i U^c_j  + \; {\rm h.c.} \; ,
\label{dil2quarksm}
 \end{equation}
exactly as in the technicolor case. When effects of the operator 
${\mathcal{O}}$ are included, this again becomes
 \begin{equation}
 m^{ij}
\left( \Delta_{\mathcal{H}} + c_q \epsilon \right)
\frac{\sigma}{f} {Q}_i U^c_j \; ,
 \end{equation}
 where $c_q$ is of order $\overline{\lambda}_{\mathcal{O}} 
f^{-\epsilon}$. In this expression 
$\Delta_{\mathcal{H}}({\hat{\lambda}}_{\mathcal{O}})$ is to be evaluated close 
to the symmetry breaking scale.

In the case where there are multiple operators $\mathcal{H}_{\alpha}$ 
that couple to the SM fermions,
 \begin{equation}
\hat{y}_U^{\alpha ij} \mathcal{H}_{\alpha} {Q}_i U^c_j  + \; {\rm h.c.} \; ,
\label{techniflavoryukawa2}
 \end{equation}
Eq.~(\ref{dil2quarksm}) generalizes to the corresponding formula in the
technicolor case, Eq.~(\ref{dil2quarks1.12}).

\subsubsection{Partially Composite Fermions}

We move on to the case where the SM quarks and leptons are partial 
composites of the strongly interacting sector. We introduce elementary 
fermions ${Q}_i, U^c_i, D^c_i, {L}_i$ and $E^c_i$ that have the same gauge 
quantum numbers as the corresponding SM fermions, and which mix with 
operators in the conformal field theory. The observed SM fermions are 
linear combinations of the corresponding elementary particles and states 
associated with the strongly interacting sector.

Mass terms for the up-type quarks arise from couplings of fermionic 
operators $\mathcal{Q}^c_{\alpha}$ and $\mathcal{U}_{\alpha}$, with 
dimensions $\Delta_{\mathcal{Q}}$ and $\Delta_{\mathcal{U}}$ 
respectively, to the elementary fermions ${Q}_i$ and $U^c_i$ in the 
Lagrangian,
 \begin{equation}
y_{\mathcal{Q}}^{\alpha i} \mathcal{Q}^c_{\alpha} Q_i +
y_{\mathcal{U}}^{\beta i} \mathcal{U}_{\beta} U^c_i  + \; {\rm h.c.}
\label{techniyukawa2.2}
 \end{equation}
We assume that the indices $\alpha$ and $\beta$, which run from 1 to 3, 
are associated with an internal U(3) symmetry of the conformal sector so 
that $\Delta_{\mathcal{Q}}$ and $\Delta_{\mathcal{U}}$ are independent 
of $\alpha$ and $\beta$. We will relax this assumption later. We can 
determine the coupling of the dilaton to the up-type quarks by promoting 
$y_{\mathcal{Q}}$ and $y_{\mathcal{U}}$ to spurions, exactly as in the 
technicolor case. Noting that $h$ has scaling dimension zero, we find 
that the conformal compensator couples as
 \begin{equation}
y^{ij} 
\left(\frac{\chi}{f} \right)^{\left(\Delta_{\mathcal{U}} + 
\Delta_{\mathcal{Q}}- 4 \right)}
\left( \hat{f} h \right) Q_{i} U^c_j  + \; {\rm h.c.} 
\label{diltocomph}
 \end{equation}
Replacing $h$ by its VEV and expanding $\chi$ out in terms of $\sigma$, 
we obtain 
 \begin{equation}
m^{ij} \left(\Delta_{\mathcal{U}} + \Delta_{\mathcal{Q}}- 4 \right)
\frac{\sigma}{f} Q_{i} U^c_j  + \; {\rm h.c.} \; ,
\label{diltocompm}
 \end{equation}
which is identical to the corresponding formula in the technicolor case, 
Eq.~(\ref{diltocomp}). When effects of the operator ${\mathcal{O}}$ are 
included, Eq.~(\ref{diltocompm}) receives corrections, and is again 
modified to
 \begin{equation}
m_{U}^{ij}
\left[ \left(\Delta_{\mathcal{U}} + \Delta_{\mathcal{Q}}- 4 \right)
+ c_q \epsilon \right]
\frac{\sigma}{f} Q_{i} U^c_j  + \; {\rm h.c.}
 \end{equation}
 In this expression, $\Delta_{\mathcal{U}}(\hat{\lambda}_{\mathcal{O}})$ 
and $\Delta_{\mathcal{Q}}(\hat{\lambda}_{\mathcal{O}})$ are to be 
evaluated close to the symmetry breaking scale. We see that corrections 
to the form of Eq.~(\ref{diltocompm}) from conformal symmetry violating 
effects are suppressed by $m_{\sigma}^2/\Lambda^2$, and are therefore 
small. In the more general case where the operators 
$\mathcal{Q}^c_{\alpha}$ and $\mathcal{U}_{\alpha}$ have dimensions 
$\Delta_{\mathcal{Q}_{\alpha}}$ and $\Delta_{\mathcal{U}_{\alpha}}$ that 
depend on the index $\alpha$, Eq.~(\ref{diltocompm}) generalizes to the 
corresponding formula in the technicolor case, 
Eq.~(\ref{diltocompquarks}).

Since $\mathcal{Q}$ and $\mathcal{U}$ are part of the strongly 
interacting sector, they must arise from complete multiplets of O(6). 
Perhaps the simplest possibility is that $\mathcal{Q}^c$ constitutes 
part of a multiplet that transforms as a fundamental of O(6), while 
$\mathcal{U}$ is just a singlet. In this realization of the extended 
symmetry $y_{\mathcal{Q}}$ violates custodial SU(2). The large mass of 
the top quark means that this coupling must be large for the third 
generation, leading to tension with precision tests. This difficulty can 
be avoided if the third generation SU(2) singlet up-type quark $U_3^c$ 
is a composite of the strongly interacting sector. This allows 
$y_{\mathcal{Q}}$ to remain small enough to avoid conflict with the 
bound. In this scenario Eq.~(\ref{diltocompquarks}) remains valid, but 
with $\Delta_{\mathcal{U}_3}$ taking the value 5/2.

\subsection{Coupling to the Higgs}

Finally we consider the dilaton coupling to the SM Higgs. In general, 
this receives contributions from both the Higgs kinetic term and the 
Higgs potential. From the kinetic term for the Higgs doublet, 
Eq.~(\ref{kinetich}), we obtain the coupling
 \begin{equation}
\frac{\sigma}{f} \partial_{\mu} \rho \partial^{\mu} \rho \; 
\label{diltohiggs1}
 \end{equation}
in the Lagrangian. Here $\rho$ is the canonically normalized SM Higgs 
field, and we are working only to quadratic order in $\rho$. When 
corrections arising from the symmetry violating parameter $\mathcal{O}$ 
are included, this is modified to
 \begin{equation}
\frac{\sigma}{f} \left[ 1 + c_H \epsilon \right]
\partial_{\mu} \rho \partial^{\mu} \rho \; .
\label{diltohiggs1.1}
 \end{equation}
where $c_H$ is of order $\overline{\lambda}_{\mathcal{O}} f^{-\epsilon}$.

The kinetic term for $\phi$, Eq.~(\ref{kineticphi}), does not lead to 
mixing between between the dilaton and the SM Higgs field. Other two
derivative terms, such as
 \begin{equation}
\frac{\partial^{\mu} \chi}{\chi} \left[ 
\phi^{\dagger} D_{\mu} \phi + (D_{\mu} \phi)^{\dagger} \phi 
\right] 
\; ,
 \end{equation}
also do not generate such mixing. This conclusion remains true when 
conformal symmetry violating effects are included.

In this scenario, the potential for the Higgs doublet can only arise 
from effects that explicitly violate the global symmetry, such as the SM 
gauge and Yukawa interactions.  If all such effects, however, respect 
conformal symmetry, then the potential for the Higgs doublet is of the very 
restrictive form
 \begin{equation}
V = \chi^4 V_0(h) \; .
\label{simpleform}
 \end{equation}
A potential of this form does not lead to mixing between the SM Higgs 
and the dilaton after minimization. The reason is that when the Higgs 
field is expanded about its VEV, there is no linear term in $\rho$ at 
the minimum of the potential $V(h)$. However, expanding $V(h)$ to 
quadratic order in $\rho$, we find a coupling of the dilaton to the 
Higgs of the form
 \begin{equation}
2 \frac{\sigma}{f} m_\rho^2 \rho^2
\label{diltohiggs2}
 \end{equation}
in the potential. This formula will receive corrections from any 
contribution to the Higgs potential that arises from an effect that 
violates conformal symmetry. Mixing between the Higgs and the dilaton 
may be generated by such effects.

In particular, when effects arising from the operator ${\mathcal{O}}$ is 
taken into account, the Higgs potential takes the more general form
 \begin{equation}
V = \chi^4 V_0(h) +
\sum_{n =1}^{\infty} \overline{\lambda}_{\mathcal{O}}^n \;
\chi^{\left(4 - n \epsilon \right)} V_n(h) \; .
 \end{equation}
Eq.~(\ref{diltohiggs2}) is consequently modified to
 \begin{equation}
\frac{\sigma}{f} \left( 2 + c_{\rho} \epsilon \right) m_\rho^2 \rho^2 \; ,
 \end{equation}
where, if the symmetry violating terms contribute significantly to the 
potential so that at the minimum $V_1(h)$ is of order $V_0(h)$, 
$c_{\rho}$ is expected to be of order $\overline{\lambda}_{\mathcal{O}} 
f^{-\epsilon}$. We see that the corrections are suppressed by 
$m_{\sigma}^2/\Lambda^2$. This effect also gives rise to mixing between 
the Higgs and the dilaton. However the mixing angle $\theta$ is 
small,
 \begin{equation}
\theta \lesssim 
\epsilon 
\overline{\lambda}_{\mathcal{O}} f^{- \epsilon} \left(\frac{v}{f}\right) 
\sim 
\frac{m_{\sigma}^2}{\Lambda^2} \left(\frac{v}{f}\right) \; .
 \end{equation}
Here $v$ is the electroweak VEV.

Since the SM gauge interactions also constitute an explicit breaking of
conformal symmetry, there will be additional radiative corrections to
the Higgs potential that are not of the simple form of
Eq.~(\ref{simpleform}). However, because the gauge interactions respect
conformal symmetry at the classical level, and only break it through
quantum effects, deviations away from this form are further loop
suppressed, and generally small.

In theories where the top quarks are elementary or partially composite,
the top Yukawa coupling also violates conformal symmetry. Then, if
contributions to the Higgs potential from loops involving the top Yukawa
coupling are sizable, 
there can be significant deviations away from
the form of Eq.~(\ref{simpleform}). We parametrize the coupling of
the conformal compensator to the top quark as
 \begin{equation}
\frac{m_t}{v} \left(\frac{\chi}{f} \right)^{(1 + \bar{\Delta})} 
\left( \hat{f} h \right) \bar{t} t \; ,
 \end{equation}
where $\bar{\Delta}$ is equal to zero if the top quarks are composite, 
is equal to $(\Delta_{\mathcal{H}} - 1)$ if the top quarks are 
elementary, and is equal to $(\Delta_{\mathcal{U}_3} + 
\Delta_{\mathcal{Q}_3}- 5)$ if the top quarks are partially composite. 
Then one loop corrections to the Higgs potential from the top loop, 
which we label by $\delta V_t$, are of the form
 \begin{equation}
\chi^4 \left[
\sum_{n =1}^{2} \frac{\hat{\alpha}_{t,n}}{\left( 16\pi^2 \right)^{n-1}}
\; \left( \frac{m_t}{v} \right)^{2n}
\left(\frac{\chi}{f}\right)^{2 n \bar{\Delta}} |h|^{2n} \right] \;.
 \end{equation}
Here the dimensionless parameters $\hat{\alpha}_{t,n}$ are of order 
one. Then Eq.~(\ref{diltohiggs2}) is modified to
 \begin{equation}
\frac{\sigma}{f} \left[ 2 + \bar{c}_{\rho} \bar{\Delta}  \right]
m_\rho^2 \rho^2 \; .
 \end{equation}
If contributions to the Higgs potential arising from loops involving the 
top Yukawa coupling are significant, so that $\delta V_t$ is comparable 
to $V$ in Eq.~(\ref{simpleform}) at the minimum, we expect 
$\bar{c}_{\rho}$ to be of order one. This effect also gives rise to 
mixing between the dilaton and the Higgs, but the mixing angle $\theta 
\lesssim \bar{\Delta} v/f$ is expected to be small in realistic models. 
Mixing will correct the dilaton couplings to other SM fields as well, 
and so a precise determination of these interactions requires this 
effect to be taken into account. We leave this for future work.

\section{Conclusions}

We have considered scenarios where strong conformal dynamics constitutes 
the ultraviolet completion of the physics responsible for electroweak 
symmetry breaking. We have constructed the effective theory of a light 
dilaton in such a framework, taking into account the explicit conformal 
symmetry violating effects that are necessarily present in realistic 
models. We have considered both the case when the corrections to the 
scaling behavior of the operator that breaks the conformal symmetry are 
small, and the case when they are large. Of particular interest is 
question whether the dilaton can naturally be light. We have shown that 
although the presence of a light dilaton is associated with tuning, the 
tuning is mild, scaling with the mass of the dilaton rather than with 
the square of the mass. As part of our analysis we have obtained the 
couplings of the dilaton to gauge bosons and fermions in the technicolor 
and Higgs as a pNGB cases, establishing results which are valid beyond 
the large $N$ limit. We have also determined the size of the corrections 
to these couplings from conformal symmetry violating effects, and found 
that they are under good theoretical control in theories where the 
dilaton is light.

\acknowledgments

We thank Daniel Stolarski for collaboration during the early stages of 
this work. We thank Roberto Franceschini and Markus Luty for useful 
discussions. We would particularly like to thank Raman Sundrum for many 
hours of invaluable discussions, and for his feedback on the manuscript. 
ZC and RKM are supported by the NSF under grant PHY-0968854.

\bigskip
\noindent
{\bf{{Note Added:}}}
While we were completing this paper, we were informed of the upcoming 
work~\cite{Bellazzini:2012vz} which overlaps with some of the ideas 
presented here.

\end{document}